\documentclass[acmsmall]{acmart}
\AtBeginDocument{%
  }

\renewcommand\footnotetextcopyrightpermission[1]{} 
\makeatletter

\usepackage{enumitem}
\usepackage{makecell}
\usepackage{graphicx}
\usepackage{subfig}
\usepackage{float}
\usepackage{booktabs}
\usepackage{fancyhdr}
\usepackage{hyperref}

\usepackage{array}
\usepackage{multirow} 

\usepackage[tikz]{bclogo}
\usepackage[framemethod=tikz]{mdframed}
\usepackage{tikz}
\usepackage[edges]{forest}
\usepackage{multicol}
\usetikzlibrary{trees,positioning,mindmap,backgrounds}
\usetikzlibrary{calc,positioning,intersections}
\usepackage{listings}
\usetikzlibrary{trees,positioning,shapes,shadows,arrows.meta}
\definecolor{hidden-draw}{RGB}{195,82,66}
\definecolor{hidden-blue}{RGB}{194,232,247}
\definecolor{hidden-orange}{RGB}{195,82,66}
\definecolor{hidden-yellow}{RGB}{242,244,193}
\definecolor{tree-level-1}{RGB}{195,82,66}
\definecolor{tree-level-2}{RGB}{195,82,66}
\definecolor{tree-level-3}{RGB}{195,82,66}
\definecolor{tree-leaf}{RGB}{195,82,66}

\usepackage{forest}
\begin{document}

\title{Towards Generalizable Human Activity Recognition: A Survey}



\author{Yize Cai}
\authornote{Both authors contributed equally to this research.}
\orcid{0009-0007-3889-474X}
\affiliation{%
  \institution{Hong Kong University of Science \& Technology (Guangzhou)}
  \country{China}
}
\email{yizecaai@gmail.com}

\author{Baoshen Guo}
\authornotemark[1]
\orcid{0000-0002-7435-8238}
\affiliation{%
  \institution{SMART, Massachusetts Institute of Technology}
  \country{Singapore}
}
\email{baoshen.guo@smart.mit.edu}

\author{Flora Salim}
\orcid{0000-0002-1237-1664}
\affiliation{%
  \institution{University of New South Wales (UNSW)}
  \country{Australia}
}
\email{flora.salim@unsw.edu.au}

\author{Zhiqing Hong}
\orcid{0000-0003-3682-4290}
\affiliation{%
  \institution{Hong Kong University of Science \& Technology (Guangzhou)}
  \country{China}
}
\email{zhiqing@berkeley.edu}




\begin{abstract}
As a critical component of Wearable AI, IMU-based Human Activity Recognition (HAR) has attracted increasing attention from both academia and industry in recent years. 
Although HAR performance has improved considerably in specific scenarios, its \textit{generalization} capability remains a key barrier to widespread real-world adoption. 
For example, domain shifts caused by variations in users, sensor positions, or environments can significantly decrease the performance in practice. 
As a result, in this survey, we explore the rapidly evolving field of IMU-based generalizable HAR, reviewing 229 research papers alongside 25 publicly available datasets to provide a broad and insightful overview. 
We first present the background and overall framework of IMU-based HAR tasks, as well as the generalization-oriented training settings. 
Then, we categorize representative methodologies from two perspectives:
(i) model-centric approaches, including pre-training method, end-to-end method, and large language model (LLM)-based learning method; and
(ii) data-centric approaches, including multi-modal learning and data augmentation techniques.
In addition, we summarize widely used datasets in this field, as well as relevant tools and benchmarks.
Building on these methodological advances, the broad applicability of IMU-based HAR is also reviewed and discussed. 
Finally, we discuss persistent challenges (e.g., data scarcity, efficient training, and reliable evaluation) and also outline future directions for HAR, including the adoption of foundation and large language models, physics-informed and context-aware reasoning, generative modeling, and resource-efficient training and inference. 
The complete list of this survey is available at \href{https://github.com/rh20624/Awesome-IMU-Sensing}{https://github.com/rh20624/Awesome-IMU-Sensing}, which will be updated continuously.

\end{abstract}






\maketitle

\section{Introduction}

\begin{figure}[t]
    \centering
    \subfloat[IMU-based HAR]{
        \includegraphics[width=2.6in]{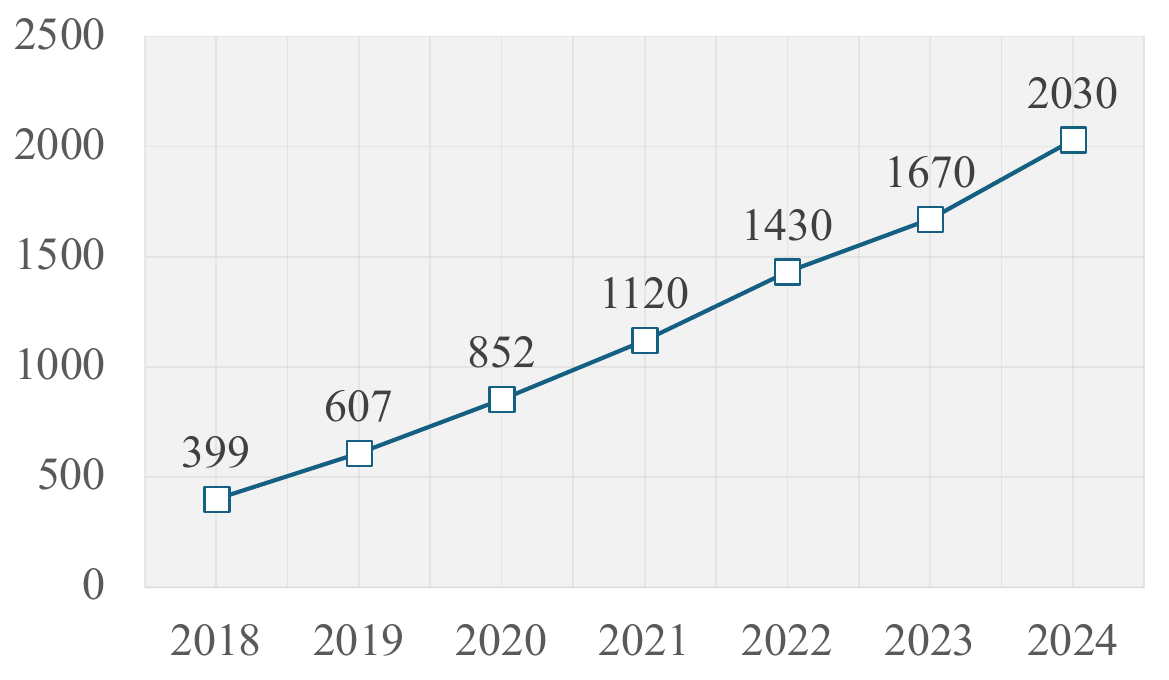}
        \label{fig:trend1}
    }\hfill
    \subfloat[Generalizable IMU-based HAR]{
        \includegraphics[width=2.6in]{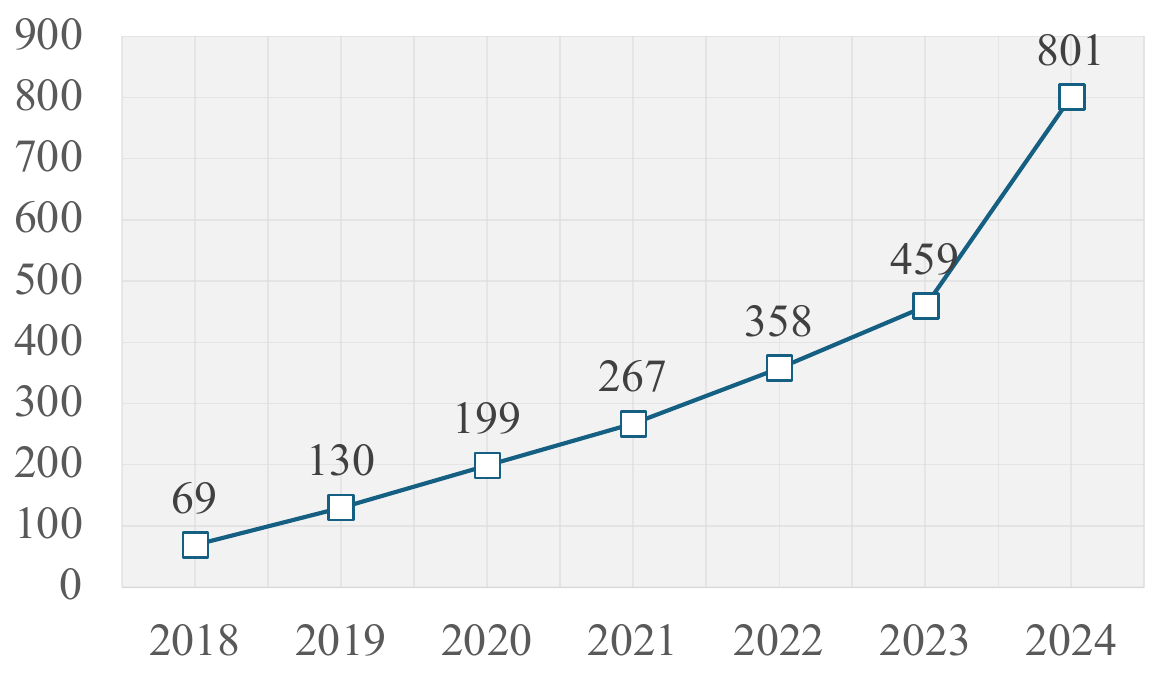}
        \label{fig:trend2}
    }
    \caption{Growth trends of cumulative publications per year (source: Google Scholar). 
    (a) Query = ``IMU'' AND ``Human Activity Recognition'', 
    (b) Query = ``IMU'' AND ``Human Activity Recognition'' AND ``Generalization''.}
    \label{fig:trend}
\end{figure}

Inertial Measurement Units (IMUs), typically comprising accelerometers, gyroscopes, and occasionally magnetometers, have become pervasive in mobile and wearable devices \cite{imwut11years}, such as smartphones, smartwatches, smart rings, smart glasses, fitness trackers, and augmented/virtual reality (AR/VR) headsets. 
The adoption of mobile and wearable technology continues to grow. 
The fitness-focused segment of wearable technology was valued at USD 14.59 billion in 2024 and is projected to reach USD 36.95 billion by 2032 \cite{Wearable2025}. 
Widely embedded in these mobile devices, IMUs produce rich time-series data that capture fine-grained human motion, making them popular for \textit{Human Activity Recognition (HAR)} in ubiquitous computing. 
HAR refers to the automatic identification of physical activities (e.g., \textit{walking}, \textit{bicycling}, \textit{sitting}) from sensor signals, enabling widespread applications such as fitness tracking, health monitoring, and human-computer interaction \cite{Kwapisz2010,plotz2011feature}. 
Driven by the advantages of IMUs, including low-cost, energy-efficient, and high penetration rate, IMU-based HAR has attracted increasing research interests over the past years. 
As shown in Fig. \ref{fig:trend1}, the number of publications in this field has increased from 399 to 2,030 in 7 years. 

\tikzstyle{my-box}= [
    rectangle,
    draw=hidden-draw,
    rounded corners,
    text opacity=1,
    minimum height=1.5em,
    minimum width=5em,
    inner sep=2pt,
    align=center,
    fill opacity=.5,
]
\tikzstyle{leaf}=[
my-box, 
minimum height=1.5em,
fill=yellow!27, 
text=black,
align=left,
font=\scriptsize,
inner xsep=5pt,
inner ysep=4pt,
align=left,
text width=45em,
]
\tikzstyle{leaf2}=[
my-box, 
minimum height=1.5em,
fill=purple!22, 
text=black,
align=left,
font=\scriptsize,
inner xsep=5pt,
inner ysep=4pt,
]
\tikzstyle{leaf3}=[
my-box, 
minimum height=1.5em,
fill=hidden-blue!57, 
text=black,
align=left,
font=\scriptsize,
inner xsep=5pt,
inner ysep=4pt,
]
\tikzstyle{leaf4}=[
my-box, 
minimum height=1.5em,
fill=green!17, 
text=black,
align=left,
font=\scriptsize,
inner xsep=5pt,
inner ysep=4pt,
]
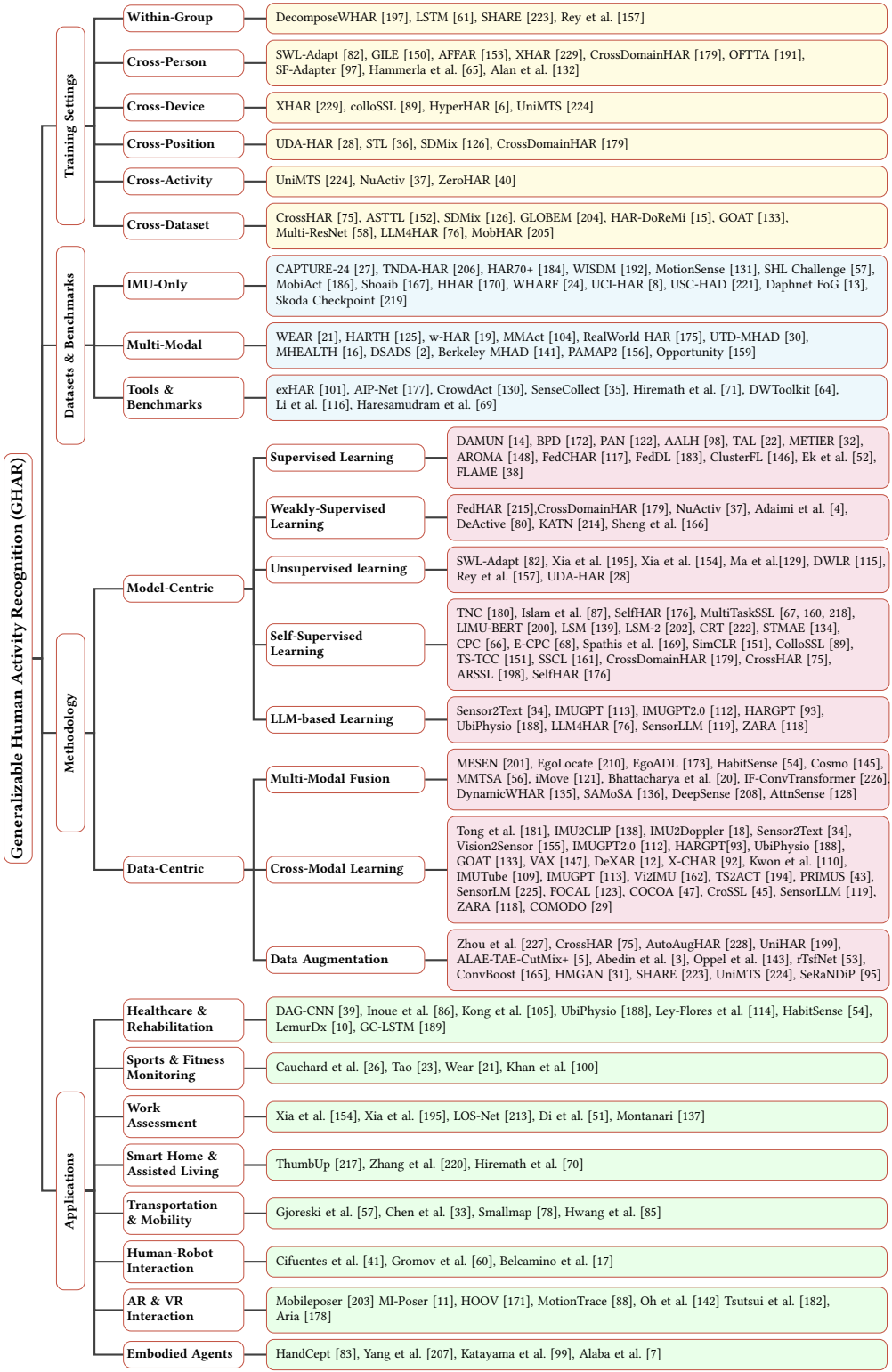
\begin{figure*}[!htbp]
\centering
\scalebox{0.8}{
  \begin{forest}
    forked edges,
    for tree={
      grow=east,
      reversed=true,
      anchor=base west,
      parent anchor=east,
      child anchor=west,
      base=left,
      font=\small,
      rectangle,
      draw=hidden-orange,
      rounded corners,
      align=left,
      minimum width=4em,
      edge+={darkgray, line width=1pt},
      s sep=3pt,
      inner xsep=2pt,
      inner ysep=3pt,
      ver/.style={rotate=90, child anchor=north, parent anchor=south, anchor=center, align=center, my-box},
    },
    where level=1{text width=10em,font=\scriptsize,}{},
    where level=2{text width=6em,font=\scriptsize,}{},
    where level=3{text width=8em,font=\scriptsize,}{},
    where level=4{text width=6.4em,font=\scriptsize,}{},
[
    \textbf{Generalizable Human Activity Recognition (GHAR)}, ver
        [
            \textbf{Training Settings}, ver
                [
                   \textbf{Within-Group}
                        [
                            DecomposeWHAR \cite{xie2025decomposing}{,}
                            LSTM \cite{guan2017ensembles}{,}
                            SHARE \cite{zhang2023unleashing}{,}
                            Rey et al. \cite{rey2017label}
                            ,leaf, text width=32em 
                        ]
                ]
                [
                    \textbf{Cross-Person}
                        [
                            SWL-Adapt \cite{cross_user_aaai23}{,}
                            GILE \cite{cross_person_aaai21}{,}
                            AFFAR \cite{cross_user_tist22}{,}
                            XHAR \cite{zhou2020xhar}{,} 
                            CrossDomainHAR~\citep{cross_user_position_TIST25}{,}
                            OFTTA \cite{wang2024optimization}{,} \\
                            SF-Adapter \cite{kang2024sf}{,} 
                            Hammerla et al. \cite{Hammerla2016}{,}
                            Alan et al. \cite{mazankiewicz2020incremental}
                            ,leaf, text width=32em
                        ]
                ]
                [
                    \textbf{Cross-Device}
                        [
                            XHAR \cite{zhou2020xhar}{,}
                            colloSSL \cite{jain2022collossl}{,}
                            HyperHAR \cite{ahmad2024hyperhar}{,}
                            UniMTS~\cite{zhang2024unimts}
                            ,leaf, text width=32em
                        ]
                ]
                [
                    \textbf{Cross-Position}
                        [
                            UDA-HAR \cite{cross_location_har_2020}{,}
                            STL \cite{cross_position_jindong_2019}{,}
                            SDMix \cite{sdmix}{,}
                            CrossDomainHAR \cite{cross_user_position_TIST25},
                            leaf, text width=32em
                        ]
                ]
                [
                    \textbf{Cross-Activity}
                        [
                            UniMTS \cite{zhang2024unimts}{,}
                            NuActiv \cite{newactiv_mobisys13}{,}
                            ZeroHAR \cite{chowdhury2025zerohar}
                            ,leaf, text width=32em
                        ]
                ]                    
                [
                    \textbf{Cross-Dataset}
                        [
                            CrossHAR \cite{crosshar}{,}
                            ASTTL \cite{ubicomp19_har_cross_dataset}{,}
                            SDMix \cite{sdmix}{,}
                            GLOBEM \cite{cross_dataset_depression}{,}
                            HAR-DoReMi \cite{cross_dataset_ban2025har}{,}
                            GOAT \cite{cross_dataset_miao2024goat}{,} \\
                            Multi-ResNet \cite{cross_dataset_ubicomp19}{,}
                            LLM4HAR \cite{LLM4HAR}{,} 
                            MobHAR \cite{xue2025mobhar},
                            leaf, text width=32em
                        ]
                ]
        ]
        [
            \textbf{Datasets \& Benchmarks}, ver
                [
                \textbf{IMU-Only}
                    [
                    CAPTURE-24~\cite{chan2024capture}{,}
                    TNDA-HAR~\cite{yan2021tndahar}{,}
                    HAR70+ ~\cite{ustad2023validation}{,}
                    WISDM~\cite{weiss2019wisdm}{,}
                    MotionSense~\cite{malekzadeh2018protecting}{,} 
                    SHL Challenge~\cite{gjoreski2018benchmarking}{,} \\
                    MobiAct~\cite{vavoulas2016mobiact}{,}
                    Shoaib~\cite{shoaib2016complex}{,}
                    HHAR~\cite{stisen2015smart}{,}
                    WHARF~\cite{bruno2013analysis}{,}
                    UCI-HAR~\cite{anguita2013public}{,} 
                    USC-HAD~\cite{mi12:ubicomp-sagaware}{,}
                    Daphnet FoG~\cite{bachlin2009wearable}{,} \\
                    Skoda Checkpoint~\cite{zappi2008activity},
                    leaf3, text width=32em
                    ]
                ]
                [
                \textbf{Multi-Modal}
                    [
                    WEAR~\cite{bock2024wear}{,}
                    HARTH~\cite{logacjov2021harth}{,}
                    w-HAR~\cite{bhat2020whar}{,}
                    MMAct~\cite{Kong_2019_ICCV}{,}
                    RealWorld HAR~\cite{sztyler2016body}{,}
                    UTD-MHAD~\cite{chen2015utd-mhad}{,} \\
                    MHEALTH~\cite{banos2014mhealth}{,}
                    DSADS~\cite{rajasegarar2013daily}{,}
                    Berkeley MHAD~\cite{Berkeley-MHAD}{,}
                    PAMAP2~\cite{reiss2012introducing}{,}
                    Opportunity~\cite{roggen2010collecting}
                    ,leaf3, text width=32em
                    ]
                ]
                [
                \textbf{Tools \&} \\ \textbf{Benchmarks} 
                    [ %
                    exHAR \cite{kianpisheh2024exhar}{,}
                    AIP-Net \cite{tanigaki2022predicting}{,}
                    CrowdAct \cite{mairittha2021crowdact}{,}
                    SenseCollect \cite{chen2021sensecollect}{,}
                    Hiremath et al. \cite{hiremath2020deriving}{,}
                    DWToolkit \cite{haladjian2019wearables}{,} \\
                    Li et al. \cite{li2017progress}{,}
                    Haresamudram et al. \cite{haresamudram2025past}
                    ,leaf3, text width=32em
                    ]
                ] 
        ]
        [
            \textbf{Methodology}, ver
                [
                \textbf{Model-Centric}
                    [
                        \textbf{Supervised Learning}
                        [
                        DAMUN \cite{bai2020adversarial}{,}
                        BPD \cite{su2022learning}{,}
                        PAN \cite{liu2019privacy}{,}
                        AALH \cite{kang2022augmented}{,}
                        TAL \cite{bock2024temporal}{,}
                        METIER \cite{chen2020metier}{,} \\
                        AROMA \cite{peng2018aroma}{,}
                        FedCHAR \cite{Chen2023FedCHAR}{,}
                        FedDL \cite{tu2021feddl}{,}
                        ClusterFL \cite{ouyang2021clusterfl}{,}
                        Ek et al. \cite{ek2020evaluation}{,} \\
                        FLAME \cite{cho2022flame}
                        ,leaf2, text width=22.5em
                        ]
                    ]
                    [
                        \textbf{Weakly-Supervised} \\ \textbf{Learning}
                        [
                        FedHAR \cite{yu2021fedhar}{,}CrossDomainHAR~\citep{cross_user_position_TIST25}{,}
                        NuActiv \cite{newactiv_mobisys13}{,}
                        Adaimi et al. \cite{adaimi2019leveraging}{,} \\
                        DeActive \cite{hossain2018deactive}{,}
                        KATN \cite{you2021katn}{,}
                        Sheng et al. \cite{sheng2020weakly}
                        ,leaf2, text width=22.5em
                        ]
                    ]
                    [
                        \textbf{Unsupervised learning}
                        [
                         SWL-Adapt \cite{cross_user_aaai23}{,}
                         Xia et al. \cite{ubicomp20_har_factory}{,}
                         Xia et al. \cite{ubicomp19_har_factory}{,}
                         Ma et al.\cite{ma2021unsupervised}{,}
                         DWLR \cite{li2024dwlr}{,} \\
                         Rey et al. \cite{rey2017label}{,}
                         UDA-HAR \cite{cross_location_har_2020}
                         ,leaf2, text width=22.5em
                        ]
                    ]
                    [
                        \textbf{Self-Supervised} \\ \textbf{Learning}
                        [
                        TNC \cite{tonekaboni2021unsupervised}{,}
                        Islam et al. \cite{islam2021self}{,}
                        SelfHAR \cite{tang2021selfhar}{,}
                        MultiTaskSSL \cite{saeed2019multi, haresamudram2022assessing, yuan2024self}{,} \\
                        LIMU-BERT \cite{Xu2021}{,}
                        LSM \cite{narayanswamy2024scaling}{,}
                        LSM-2 \cite{xu2025lsm}{,}
                        CRT \cite{zhang2023self}{,}
                        STMAE \cite{miao2024spatial}{,} \\
                        CPC \cite{haresamudram2021contrastive}{,}
                        E-CPC \cite{haresamudram2023investigating}{,} 
                        Spathis et al. \cite{spathis2020learning}{,} 
                        SimCLR \cite{qian2022makes}{,}
                        ColloSSL \cite{jain2022collossl}{,} \\
                        TS-TCC \cite{qian2022makes}{,}
                        SSCL \cite{saeed2020federated}{,}
                        CrossDomainHAR~\citep{cross_user_position_TIST25}{,} 
                        CrossHAR \cite{crosshar}{,} \\
                        ARSSL \cite{xu2023augmentation}{,}
                        SelfHAR \cite{tang2021selfhar}
                        ,leaf2, text width=22.5em
                        ]
                    ]
                    [
                        \textbf{LLM-based Learning}
                        [
                        Sensor2Text \cite{chen2024sensor2text}{,}
                        IMUGPT \cite{leng2023generating}{,} 
                        IMUGPT2.0 \cite{leng2024imugpt}{,}
                        HARGPT \cite{ji2024hargpt}{,} \\
                        UbiPhysio \cite{wang2024ubiphysio}{,}
                        LLM4HAR \cite{LLM4HAR}{,}
                        SensorLLM \cite{li2024sensorllm}{,} %
                        ZARA \cite{li2025zara} %
                        ,leaf2, text width=22.5em
                        ]
                    ]
                ]
                [
                \textbf{Data-Centric}
                    [
                        \textbf{Multi-Modal Fusion}
                        [
                        MESEN \cite{xu2023mesen}{,}
                        EgoLocate \cite{yi2023egolocate}{,}
                        EgoADL \cite{sun2024multimodal}{,}
                        HabitSense \cite{fernandes2024habitsense}{,} 
                        Cosmo \cite{ouyang2022cosmo}{,} \\ 
                        MMTSA \cite{gao2023mmtsa}{,} 
                        iMove \cite{liu2024imove}{,} 
                        Bhattacharya et al. \cite{bhattacharya2022leveraging}{,} 
                        IF-ConvTransformer \cite{zhang2022if}{,} \\ 
                        DynamicWHAR \cite{miao2022towards}{,}
                        SAMoSA \cite{mollyn2022samosa}{,}
                        DeepSense \cite{Yao2017}{,}
                        AttnSense \cite{Ma2019}
                        ,leaf2, text width=22.5em
                        ]
                    ]
                    [
                        \textbf{Cross-Modal Learning}
                        [
                        Tong et al. \cite{tongzeroshot21}{,}
                        IMU2CLIP \cite{moon2023imu2clip}{,}
                        IMU2Doppler \cite{bhalla2021imu2doppler}{,}
                        Sensor2Text \cite{chen2024sensor2text}{,} \\
                        Vision2Sensor \cite{radu2019vision2sensor}{,} 
                        IMUGPT2.0 \cite{leng2024imugpt}{,}
                        HARGPT\cite{ji2024hargpt}{,}  %
                        UbiPhysio \cite{wang2024ubiphysio}{,} \\
                        GOAT \cite{cross_dataset_miao2024goat}{,}
                        VAX \cite{patidar2023vax}{,} 
                        DeXAR \cite{arrotta2022dexar}{,}
                        X-CHAR \cite{jeyakumar2023x}{,}
                        Kwon et al. \cite{kwon2021approaching}{,} \\
                        IMUTube \cite{Kwon2020}{,} 
                        IMUGPT \cite{leng2023generating}{,} 
                        Vi2IMU \cite{santhalingam2023synthetic}{,}
                        TS2ACT \cite{xia2024ts2act}{,}
                        PRIMUS \cite{das2024primus}{,} \\
                        SensorLM \cite{zhang2025sensorlm}{,}
                        FOCAL \cite{liu2023focal}{,}
                        COCOA \cite{deldari2022cocoa}{,}
                        CroSSL \cite{deldari2024crossl}{,}
                        SensorLLM \cite{li2024sensorllm}{,} \\
                        ZARA \cite{li2025zara}{,}
                        COMODO \cite{chen2025comodo}
                        ,leaf2, text width=22.5em
                        ]
                    ]
                    [
                        \textbf{Data Augmentation}
                        [
                        Zhou et al. \cite{zhou2025learning}{,}
                        CrossHAR \cite{crosshar}{,}
                        AutoAugHAR \cite{zhou2024autoaughar}{,} 
                        UniHAR \cite{unihar}{,} \\
                        ALAE-TAE-CutMix+ \cite{ahmad2023alae}{,}
                        Abedin et al. \cite{abedin2021attend}{,} 
                        Oppel et al. \cite{oppel2025diffusion}{,} 
                        rTsfNet \cite{enokibori2024rtsfnet}{,} \\ 
                        ConvBoost \cite{shao2023convboost}{,}  
                        HMGAN \cite{chen2023hmgan}{,} 
                        SHARE \cite{zhang2023unleashing}{,} 
                        UniMTS \cite{zhang2024unimts}{,}  
                        SeRaNDiP \cite{Kalupahana2023SeRaNDiP} 
                        ,leaf2, text width=22.5em
                        ]
                    ]
                ]
        ]
        [
            \textbf{Applications}, ver
                [
                \textbf{Healthcare \&} \\ \textbf{Rehabilitation}
                    [
                    DAG-CNN \cite{choi2022deep}{,}
                    Inoue et al. \cite{inoue2019integrating}{,}
                    Kong et al. \cite{kong2013development}{,} %
                    UbiPhysio \cite{wang2024ubiphysio}{,}
                    Ley-Flores et al. \cite{ley2024co}{,} 
                    HabitSense \cite{fernandes2024habitsense}{,} \\
                    LemurDx \cite{arakawa2023lemurdx}{,}
                    GC-LSTM \cite{wang2021leveraging}
                    ,leaf4, text width=32em
                    ]
                ]
                [
                \textbf{Sports \& Fitness} \\ \textbf{Monitoring}
                    [
                    Cauchard et al. \cite{cauchard2019positive}{,}
                    Tao \cite{boovaraghavan2023tao}{,}
                    Wear \cite{bock2024wear}{,} %
                    Khan et al. \cite{khan2017activity}
                    ,leaf4, text width=32em
                    ]
                ]
                [
                \textbf{Work} \\ \textbf{Assessment}
                    [
                    Xia et al. \cite{ubicomp19_har_factory}{,}
                    Xia et al. \cite{ubicomp20_har_factory}{,}
                    LOS-Net \cite{yoshimura2022acceleration}{,}
                    Di et al. \cite{di2020multi}{,}
                    Montanari \cite{montanari2017detecting}
                    ,leaf4, text width=32em
                    ]
                ] 
                [
                \textbf{Smart Home \&} \\ \textbf{Assisted Living}  
                    [
                    ThumbUp \cite{yu2022thumbup}{,}
                    Zhang et al. \cite{zhang2021fine}{,} %
                    Hiremath et al. \cite{hiremath2022bootstrapping}
                    ,leaf4, text width=32em
                    ]
                ]  
                [
                \textbf{Transportation} \\ \textbf{\& Mobility}
                    [
                    Gjoreski et al. \cite{gjoreski2018benchmarking}{,}
                    Chen et al. \cite{chen2023enhancing}{,}
                    Smallmap \cite{hong2024smallmap}{,}
                    Hwang et al. \cite{hwang2024more}
                    ,leaf4, text width=32em
                    ]
                ] 
                [
                \textbf{Human-Robot} \\ \textbf{Interaction}
                    [
                    Cifuentes et al. \cite{cifuentes2014human}{,}
                    Gromov et al. \cite{gromov2019proximity}{,}
                    Belcamino et al. \cite{belcamino2025distributed},
                    leaf4, text width=32em
                    ]
                ]
                [
                \textbf{AR \& VR} \\ \textbf{Interaction}
                    [
                    Mobileposer \cite{xu2024mobileposer}
                    MI-Poser \cite{arakawa2023mi}{,}
                    HOOV \cite{streli2023hoov}{,}
                    MotionTrace \cite{islam2024motiontrace}{,}
                    Oh et al. \cite{oh2024uwb}{}
                    Tsutsui et al. \cite{tsutsui2021you}{,} \\
                    Aria \cite{aria2023dataset},
                    leaf4, text width=32em
                    ]
                ]
                [
                \textbf{Embodied Agents }
                    [
                    HandCept \cite{huang2025handcept}{,}
                    Yang et al. \cite{yang2023multi}{,}
                    Katayama et al. \cite{katayama2025learning}{,}
                    Alaba et al. \cite{alaba2024gps},
                    leaf4, text width=32em
                    ]
                ]    
        ]
]
    \end{forest}
    }
    \vspace{-0.1in}
    \caption{Framework of this survey.}
    \label{fig:survey}
\end{figure*}

In existing HAR studies, the earliest attempts relied on classical machine learning approaches (e.g., shallow classifiers such as decision trees and SVMs~\cite{har_ml_2011}) with hand-crafted features from IMU signals (e.g., mean, variance, fast Fourier transform (FFT) coefficients). 
While effective for simple activities, these methods were constrained by manual feature engineering and struggled with complex or subtle motion patterns. 
In recent years, the emergence of deep learning marked a turning point. Deep Models like convolutional neural networks (CNNs)~\cite{krizhevsky2012imagenet}, recurrent neural networks (RNNs)~\cite{hochreiter1997long}, and transformer~\cite{vaswani2017attention} enabled automatic extraction of discriminative features directly from raw sensor data. 
Pioneering work~\cite{Ordonez2016} introduces a CNN-LSTM architecture that significantly outperformed traditional models on multimodal wearable datasets. 
Similarly, Hammerla et al.~\cite{Hammerla2016} conduct a rigorous comparison between deep and shallow models, demonstrating that deep architectures achieve superior performance. 
Furthermore, emerging paradigms such as self-supervised learning (SSL)~\cite{logacjov2024self, deldari2022beyond}, federated on-device learning~\cite{aouedi2024federated}, and multi-modal or cross-domain data fusion~\cite{ni2024survey} have also been introduced to further improve the performance of HAR tasks.

Despite significant progress, IMU-based HAR still faces fundamental challenges, particularly in \textbf{achieving generalizability across heterogeneous conditions}~\cite{xia2015realtime,reyes2016transition,kroger2019privacy}. 
According to previous studies, recognition accuracy can drop by \textbf{42.4\%} when faced with heterogeneous domain shifts~\cite{crosshar}. 
Firstly, Human motion patterns vary widely due to differences in users, sensor placements, and activity execution styles, resulting in \textbf{domain shifts} that degrade model performance when deployed across devices, scenarios, or datasets~\cite{xia2015realtime, reyes2016transition}. 
Secondly, label scarcity further exacerbates this issue, as annotating data for every new domain is costly and often impractical~\cite{logacjov2024self}. 
These factors highlight the urgent need for methods that can generalize effectively to unseen conditions. 

As shown in Fig. \ref{fig:trend2}, considering the challenges of domain shifts and the inherent difficulty of achieving generalizability in IMU-based HAR tasks, recent research trends have increasingly focused on generalizable learning paradigms.
From a data-centric perspective, multi-modal approaches integrate complementary data modalities with IMU signals or transfer knowledge from models trained on other modalities to strengthen IMU-based HAR and improve generalization. 
Additional strategies involve applying advanced data augmentation techniques to enhance data diversity and quality.
From a model-centric perspective, many advanced supervised, weakly-supervised, unsupervised, self-supervised, and LLM-based approaches aim to enhance the generalization of IMU-based HAR models and reduce reliance on labeled data.
This growing emphasis is reflected in the sharp increase in publications on sensor data generalization across both AI and data mining venues (e.g., KDD, AAAI, IJCAI, ICML, NeurIPS, TPAMI, TKDE) and ubiquitous computing venues (e.g., UbiComp/IMWUT, MobiCom, PerCom, SenSys, TMC). 
The total number of papers rose from 69 in 2018 to 801 in 2024, increasing more than 11 times over six years. 

In this survey, we focus on the \textbf{generalization capability} of IMU-based HAR models, a critical challenge for achieving robust performance in real-world applications. 
We review more than 200 research papers alongside 25 publicly available datasets, 
as shown in Fig.~\ref{fig:survey}, aiming to provide a comprehensive overview of how generalization has been addressed in the literature. 
Particularly, this survey begins with an introduction to the HAR task and its general framework, establishing a foundational context. 
We then define six distinct training settings from a generalization perspective and categorize existing approaches based on model- and data-centric strategies. 
This is followed by a thorough review of widely used datasets and a summary of key application domains in IMU-based HAR. 
Finally, we discuss the current limitations in the field and propose promising avenues for future research. 
We anticipate that this survey will serve as a valuable and practical resource for both newcomers and seasoned researchers interested in understanding, implementing, or advancing generalization techniques in IMU-based HAR. 

Table~\ref{tab:related-survey} shows the comparison between our survey and existing representative IMU-sensing-based surveys \cite{Bulling2014, Kamboj2024crossmodal, susarla2024non, logacjov2024self, Dhekane2024, ye2024machine, ni2024survey, huang2024survey, saraf2023survey, Delgado2022privacy}. 
While prior surveys primarily focus on specific individual scenarios, our work reviews five generalizable scenarios. 
From a data-centric perspective, we summarize data augmentation and multi/cross-modal techniques. 
From a model-centric perspective, unlike existing surveys that are limited to SSL, multi-task learning, or transfer learning, we conduct an extensive and in-depth review covering pre-training and fine-tuning methods, training from scratch methods, and LLM-based methods.
This provides a comprehensive overview of generalizable IMU-based HAR.

\begin{table}[h]
\centering
\caption{Comparison with existing surveys.}
\label{tab:related-survey}
\renewcommand{\arraystretch}{1.2}
\resizebox{0.95\textwidth}{!}{
\begin{tabular}{l|c|ccccc|cc|ccc}
\hline
\multirow{2}{*}{\textbf{Related Surveys}}  
& \multirow{2}{*}{\textbf{Years}}  
& \multicolumn{5}{c|}{\textbf{Generalizable Settings}} 
& \multicolumn{2}{c|}{\textbf{Data Centric}} 
& \multicolumn{3}{c}{\textbf{Model Centric}} \\ \cline{3-12}
&
& \makecell{Cross \\Person} 
& \makecell{Cross \\Device} 
& \makecell{Cross \\Position} 
& \makecell{Cross \\Activity} 
& \makecell{Cross \\Dataset}
& \makecell{Data \\Augmentation} &  \makecell{Multi \\modality}
& \makecell{Pre-Training \& \\Fine-Tuning} & \makecell{Training from \\ Scratch} & \makecell{Leveraging \\ LLM} \\ \hline
Bulling et al~\cite{Bulling2014} & 2014 & \checkmark & & & & & & & & \checkmark &  \\ 
Delgado et al~\cite{Delgado2022privacy} & 2022 &  &  & &  &  &  & & & \checkmark &  \\ 
Saraf et al~\cite{saraf2023survey} & 2023 &  &  & &  & \checkmark &  & & & \checkmark &  \\ 
Kamboj et al.~\cite{Kamboj2024crossmodal} & 2024 & \checkmark & \checkmark & \checkmark & & & & \checkmark & & \checkmark & \\ 
Susarla et al.~\cite{susarla2024non} & 2024 &  & & &  &  & \checkmark & \checkmark  & &\checkmark &  \\ 
Logacjov et al.~\cite{logacjov2024self} & 2024 & \checkmark & &  & & \checkmark & & \checkmark & \checkmark & &  \\ 
Dhekane et al~\cite{Dhekane2024} & 2024 & \checkmark &  & &  & \checkmark & & \checkmark & \checkmark  & \checkmark &  \\ 
Ye et al~\cite{ye2024machine} & 2024 & \checkmark & \checkmark & &  &  &  & \checkmark & & \checkmark &  \\ 
Ni et al~\cite{ni2024survey} & 2024 &  &  & \checkmark & \checkmark & \checkmark &  & \checkmark & \checkmark & \checkmark &  \\ 
Huang et al~\cite{huang2024survey} & 2024 &  &  & &  &  & \checkmark  & \checkmark & \checkmark & \checkmark &  \\ \hline
\textbf{Our work} & \textbf{2025} & \textbf{\checkmark} & \textbf{\checkmark} & \textbf{\checkmark} & \textbf{\checkmark} & \textbf{\checkmark} 
& \textbf{\checkmark} & \textbf{\checkmark} 
& \textbf{\checkmark} & \textbf{\checkmark} & \textbf{\checkmark} \\ \hline

\end{tabular}
}
\end{table}

The rest of the survey is structured as follows.
Section \ref{sec:background} provides background information on IMU-based HAR.
Section \ref{sec:survey} presents the survey, covering representative works across training settings, datasets, methodology, and applications.
Section \ref{sec:challenge} discusses the open challenges and proposed future directions.
Finally, Section \ref{sec:conclusion} concludes the paper.
\section{Background}
\label{sec:background}

\subsection{IMU-based HAR}

Human Activity Recognition (HAR) aims to automatically identify human physical activities from sensor data. 
Among various sensing modalities, Inertial Measurement Units (IMUs), typically composed of accelerometers, gyroscopes, and magnetometers, have become increasingly popular due to their portability, cost-effectiveness, and non-intrusive nature. 
IMU sensors measure acceleration, angular velocity, and magnetic field strength along three orthogonal axes (denoted by $x$, $y$, and $z$), providing rich temporal signals for HAR. 

Formally, given a sequence of IMU sensor readings, the HAR task can be defined as follows. 
Let $\mathbf{X}$ represent an IMU sensor data sequence:
\begin{equation}
    \mathbf{X} = [\mathbf{x}_1, \mathbf{x}_2, \dots, \mathbf{x}_T]^\top \in \mathbb{R}^{T \times D}
\end{equation}
where $T$ denotes the sequence length (number of time steps), and $D$ denotes the dimensionality of sensor readings (e.g., accelerometer and gyroscope data along three axes result in $D=6$). 
Each $\mathbf{x}_t \in \mathbb{R}^{D}$ is a vector of IMU measurements at time step $t$.
The goal of HAR is to learn a mapping function $f(\cdot)$ that predicts the activity label $y$ from the sensor sequence $\mathbf{X}$:
\begin{equation}
    f: \mathbf{X} \mapsto y, \quad y \in \mathcal{Y}
\end{equation}
where $\mathcal{Y} = \{y_1, y_2, \dots, y_C\}$ is the set of $C$ possible activity classes such as running and going upstairs.

\subsection{Framework}

\begin{figure}[t]
    \centering
    \includegraphics[width=5.5in]{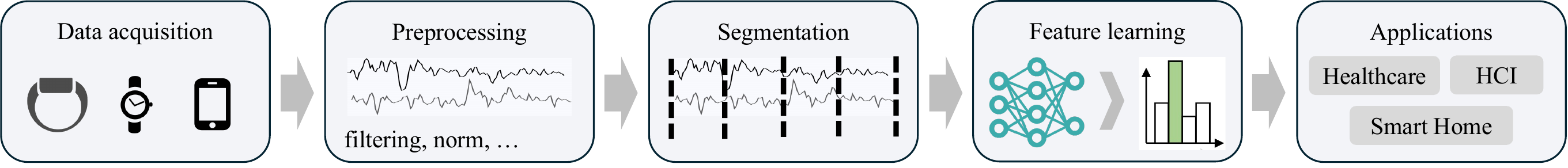}
    \caption{Illustration of a typical IMU-based HAR framework.}
    \label{fig:framework}
    \vspace{-0.2in}
\end{figure}

The development of an effective IMU-based HAR system generally follows a multi-stage processing pipeline designed to transform raw inertial signals into meaningful activity labels, as shown in Fig.~\ref{fig:framework}. 
The following outlines the five steps most commonly adopted in the literatures:

\begin{itemize}

    \item \textbf{Data acquisition:}
    Raw inertial signals are collected from wearable IMU sensors as users perform a range of predefined activities.
    
    \item \textbf{Preprocessing:}
    The acquired signals are subjected to various preprocessing techniques, including filtering to eliminate high-frequency noise, artifact removal, and normalization to ensure consistency across sensors.
    
    \item \textbf{Segmentation:}
    Continuous sensor data streams are partitioned into either fixed-length time windows or activity-specific segments to facilitate subsequent analysis and modeling.
    
    \item \textbf{Feature learning:} 
    Extract handcrafted or learned features from segments and train machine learning or deep learning models to predict activity labels.

    \item \textbf{Applications:}
    HAR models can be applied in many fields, such as healthcare for monitoring, human-computer interaction for intuitive interfaces, and smart city for map sensing.
    
\end{itemize}

The above pipeline provides a general framework for IMU-based HAR, with variations depending on the specific goals (e.g., real-time inference, domain generalization, or multi-sensor fusion). 
Each stage provides opportunities for methodological advancements, particularly in enhancing robustness under domain shifts, reducing labeling cost, or improving computational efficiency on edge devices.
\section{Generalizable IMU-based HAR}
\label{sec:survey}

Fig.~\ref{fig:survey} outlines the overall framework of this survey along with representative works that exemplify key research directions in IMU-based HAR. 
To provide a structured overview, Section \ref{sec:train} categorizes existing training settings from the perspective of generalization.
Section~\ref{sec:model} and Section~\ref{sec:data} introduce existing IMU-based methodologies from both model-centric and data-centric perspectives.
Section \ref{sec:datasets} summarizes widely used benchmarks and datasets.
Finally, Section \ref{sec:application} reviews the applications of IMU-based HAR.

\subsection{Generalization-Oriented Training Settings}
\label{sec:train}

Limited generalization capability is a major issue in IMU-based HAR, as models trained on data from specific subjects, devices, sensor placements, or datasets often suffer significant performance degradation when applied to unseen scenarios with distributional shifts. 
To mitigate this limitation, recent works have explored a variety of generalization settings designed to enhance model robustness against data heterogeneity.
In this work, we provide a systematic analysis of this issue by categorizing existing HAR tasks according to the configurations of their training and testing data.

\begin{enumerate}[leftmargin=*]

    \item \textbf{Within-group HAR (Train and test on the same group):} 
    The training and testing data are drawn from the same distribution, meaning they involve the same subjects, sensor devices, sensor positions, and dataset configuration \cite{xie2025decomposing, guan2017ensembles, zhang2023unleashing, rey2017label}.
    Formally, given a training set $\mathcal{D}_{train}$ and testing set $\mathcal{D}_{test}$, we have:
    \begin{equation}
        p(\mathbf{X}, y)_{train} \approx p(\mathbf{X}, y)_{test}
    \end{equation}
    This setting represents the simplest scenario and usually yields high accuracy, but does not reflect real-world deployment scenarios.

    \item \textbf{Cross-person HAR:} 
    Training and testing data are collected from different groups of subjects \cite{cross_user_aaai23, cross_person_aaai21, cross_user_tist22, zhou2020xhar, cross_user_position_TIST25, wang2024optimization, kang2024sf, Hammerla2016, mazankiewicz2020incremental}, introducing variability in individual movement patterns and personal characteristics. 
    Let $U_{train}$ and $U_{test}$ denote distinct sets of users, then:
    \begin{equation}
        U_{train} \cap U_{test} = \emptyset, \quad p(\mathbf{X}, y | U_{train}) \neq p(\mathbf{X}, y | U_{test})
    \end{equation}
    This scenario evaluates the model's ability to generalize across different individuals.

    \item \textbf{Cross-device HAR:} 
    Training and testing data are collected using different IMU sensor devices \cite{zhou2020xhar, jain2022collossl, ahmad2024hyperhar, zhang2024unimts}, which may differ in sampling rates, sensitivity, and noise characteristics. 
    This variation poses challenges for model generalization across hardware platforms.
    Denote $D_{train}$ and $D_{test}$ as distinct sets of devices, then:
    \begin{equation}
        D_{train} \cap D_{test} = \emptyset, \quad p(\mathbf{X}, y | D_{train}) \neq p(\mathbf{X}, y | D_{test})
    \end{equation}
    This scenario assesses the robustness of models against device heterogeneity.

    \item \textbf{Cross-position HAR:} 
    Training and testing data are collected from IMU sensors placed at different body locations (e.g., wrist, waist, ankle) \cite{cross_location_har_2020, cross_position_jindong_2019, sdmix, cross_user_position_TIST25}, introducing variability due to position-dependent motion characteristics. 
    Let $P_{train}$ and $P_{test}$ denote different sensor positions, then:
    \begin{equation}
        P_{train} \cap P_{test} = \emptyset, \quad p(\mathbf{X}, y | P_{train}) \neq p(\mathbf{X}, y | P_{test})
    \end{equation}
    This scenario tests the model's ability to generalize across varying sensor placements.

    \item \textbf{Cross-activity HAR:} 
    The test data includes activities that are not present in the training data \cite{zhang2024unimts, newactiv_mobisys13, chowdhury2025zerohar}, requiring the model to recognize previously unseen activities by leveraging knowledge learned from seen ones. 
    Let $A_{train}$ and $A_{test}$ denote different sets of activities, then:
    \begin{equation}
        A_{train} \cap A_{test} = \emptyset, \quad p(\mathbf{X}, y | A_{train}) \neq p(\mathbf{X}, y | A_{test})
    \end{equation}
    This scenario evaluates the model's ability to generalize and adapt to new activities, such as training on walking and standing while recognizing running as a new activity.

    \item \textbf{Cross-dataset HAR:} 
    This setting involves training and testing on data from distinct datasets \cite{crosshar, ubicomp19_har_cross_dataset, sdmix, cross_dataset_depression, cross_dataset_ban2025har, cross_dataset_miao2024goat, cross_dataset_ubicomp19, LLM4HAR, xue2025mobhar}, often collected under varying protocols, sensor setups, environmental contexts, and annotation methodologies, thereby posing significant challenges for model generalization and robustness. Let $\mathcal{D}^A$ and $\mathcal{D}^B$ be two distinct datasets, then:
    \begin{equation}
        \mathcal{D}_{train} \subseteq \mathcal{D}^A, \quad \mathcal{D}_{test} \subseteq \mathcal{D}^B, \quad p(\mathbf{X}, y | \mathcal{D}^A) \neq p(\mathbf{X}, y | \mathcal{D}^B)
    \end{equation}
    This scenario represents the most challenging setting, evaluating the capability of models to generalize across entirely different data collection conditions.
\end{enumerate}

Various evaluation settings in HAR reflect different levels of distributional shift between training and testing data. 
These settings progressively increase the difficulty of generalization and are critical for assessing the robustness and real-world applicability of HAR models. 

\subsection{Model-Centric Methodology}
\label{sec:model}

Methodological advancements are crucial for improving the adaptability and robustness of IMU-based HAR systems in diverse real-world settings. 
To enhance generalizability, a range of model-centric strategies has been proposed. 
Here, we categorize these approaches into supervised learning, semi-supervised learning, weakly-supervised learning, self-supervised Learning, and LLM-based Learning. 
Their primary goals are to increase robustness under domain shifts or to reduce dependence on large-scale labeled data.

\subsubsection{\textbf{Supervised Learning}}

Supervised learning has long been the predominant paradigm in HAR, relying on large-scale labeled datasets to train models that map raw sensor signals to predefined activity categories. 
Existing supervised learning methods enhance generalizability mainly through feature disentanglement, multi-task learning, and federated learning.

\begin{figure*}[t]
    \centering
    \subfloat[Feature Disentangle]{
        \centering
        \includegraphics[width=1.1in]{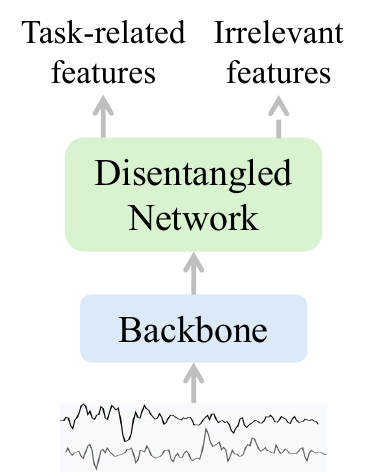}
        \label{fig:dis}
    }
    \hspace{1.0cm}
    \subfloat[Multi-Task Learning]{
        \centering
        \includegraphics[width=1.1in]{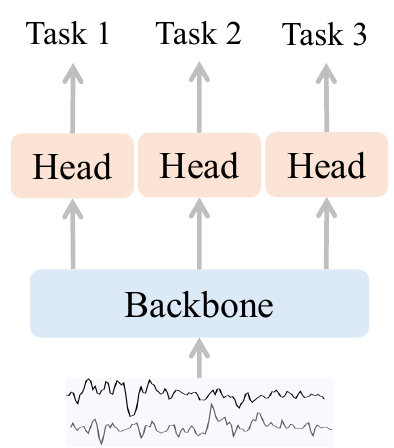}
        \label{fig:multi}
    }
    \hspace{1.0cm}
    \subfloat[Federated Learning]{
        \centering
        \includegraphics[width=1.3in]{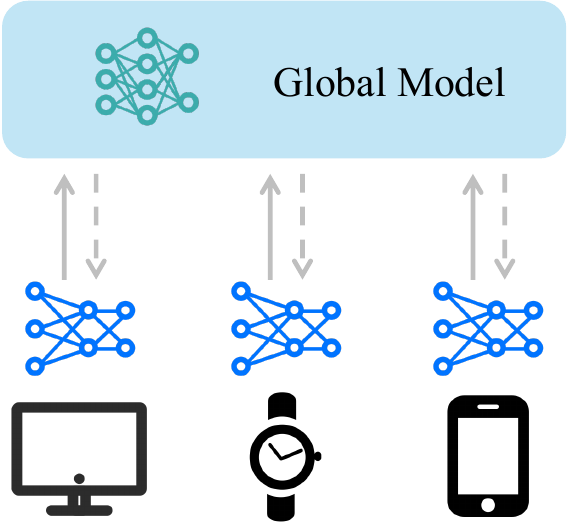}
        \label{fig:fed}
    }
    \caption{Illustration of feature disentangle, multi-task learning, and federated learning.}
    \label{fig:sup}
    \vspace{-0.1in}
\end{figure*}

\textbf{Feature Disentanglement:}
IMU signals often contain entangled information irrelevant to HAR, such as variations from device placement or individual movement styles. 
These factors introduce data heterogeneity and hinder generalization. 
Feature disentanglement seeks to separate nuisance factors from task-relevant representations, preserving only the information essential for accurate HAR, as shown in Fig. \ref{fig:dis}.
DAMUN \cite{bai2020adversarial} employs a multi-view feature extractor and fusion module to capture comprehensive representations, followed by a twin adversarial network to isolate subject-specific variations. 
BPD \cite{su2022learning} uses adversarial training to disentangle behavior patterns from nuisance factors such as individual styles and environmental noise. 
From a privacy perspective, PAN \cite{liu2019privacy} separates sensitive personal information from task-relevant features via adversarial learning, balancing privacy and generalization. 
AALH \cite{kang2022augmented} extends this by mapping data from different sensor configurations into a shared latent space, addressing variability in sensor placement and user characteristics.

\textbf{Multi-Task Learning:}
Multi-task learning jointly trains multiple supervised learning tasks, often improving performance relative to independent task learning. 
As illustrated in Fig. \ref{fig:multi}, it enhances generalization by leveraging shared representations across related tasks.
TAL \cite{bock2024temporal} integrates activity localization with HAR to boost performance. 
METIER \cite{chen2020metier} jointly learns HAR and user recognition via soft parameter sharing, while AROMA \cite{peng2018aroma} separates simple and complex HAR using a deep multi-task framework. 

\textbf{Federated Learning:}
Some approaches integrate supervised learning with federated deployment to enhance generalization.
As shown in Fig. \ref{fig:fed},federated learning is a distributed learning paradigm where clients train local models on loacal data and share only model updates with a central server. 
The server aggregates these updates to form a global model, which is redistributed to clients. 
In IMU-based HAR, federated learning facilitates generalization across heterogeneous devices. 
FedCHAR \cite{Chen2023FedCHAR} and ClusterFL \cite{ouyang2021clusterfl} integrate federated learning with client clustering to achieve more stable and personalized HAR performance. 
FedDL \cite{tu2021feddl} adopts a dynamic model sharing strategy to aggregate local supervised models, maintaining generalization across heterogeneous clients. 
Ek et al. \cite{ek2020evaluation} conduct a comprehensive evaluation of federated learning for HAR tasks. 
FLAME \cite{cho2022flame} introduces a class-based data partitioning strategy to adapt existing HAR datasets for federated learning scenarios.

\subsubsection{\textbf{Weakly Supervised Learning}}

Weakly supervised learning addresses scenarios in which the supervision signal is limited in fidelity, coverage, or granularity. 
The goal is to recover reliable, fine-grained predictions by integrating weak labels with inductive biases, structural constraints, and auxiliary learning signals.
In context of HAR, such method mainly involves semi-supervised learning, active learning, and inexact supervised learning.

\textbf{Semi-Supervised Learning:}
Semi-supervised learning aims to mitigate the heavy reliance on large-scale annotated datasets by jointly leveraging labeled data together with unlabeled data. 
The central assumption is that unlabeled data, though lacking explicit supervision, encodes structural regularities that can be exploited to improve representation learning and decision boundaries.
CrossDomainHAR \cite{cross_user_position_TIST25} leverages labeled HAR datasets to guide models in bridging the conceptual gap between source and target domains. 
FedHAR \cite{yu2021fedhar} combines semi-supervised learning with federated learning to address label scarcity, enhance generalization, and preserve privacy in distributed HAR scenarios.

\begin{figure*}[t]
    \centering
    \subfloat[Active Learning]{
        \centering
        \includegraphics[width=1.9in]{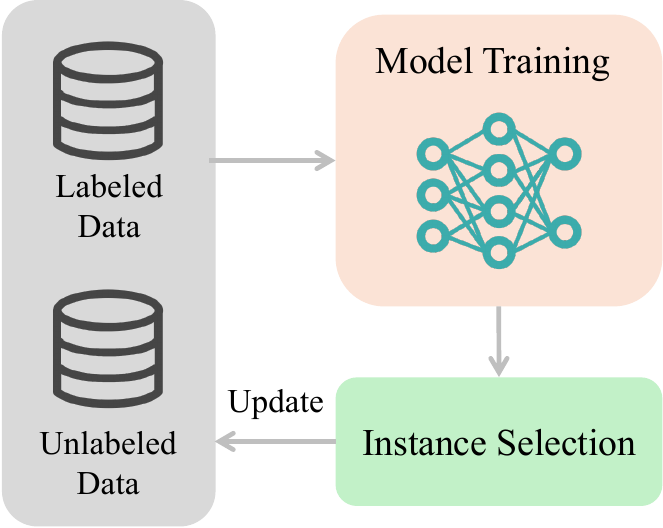}
        \label{fig:act}
    }
    \hspace{2.0cm}
    \subfloat[Siamese Network]{
        \centering
        \includegraphics[width=1.9in]{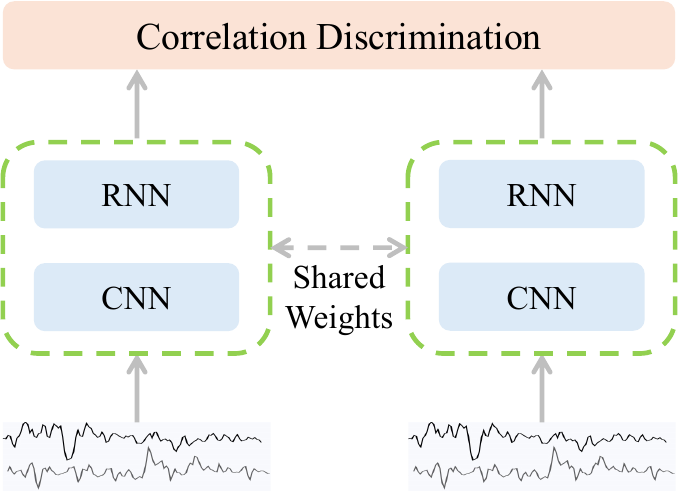}
        \label{fig:sia}
    }
    \caption{Illustration of active learning and siamese network.}
    \label{fig:sup}
    \vspace{-0.1in}
\end{figure*}

\textbf{Active Learning:}
Active Learning recognizes that not all data samples contribute equally during training. 
As shown in Fig \ref{fig:act}, given a labeled dataset and a pool of unlabeled data, the model actively selects the most informative unlabeled samples to be labeled. 
This selective annotation process aims to maximize generalizability while minimizing labeling effort, making it particularly useful for HAR tasks where annotation is costly and time-consuming.
NuActiv \cite{newactiv_mobisys13} introduces a novel activity representation based on semantic attributes and integrates it with AL to support the discovery of new activity classes. 
Adaimi et al. \cite{adaimi2019leveraging} demonstrate that AL can serve as an effective labeling strategy, achieving performance comparable to or exceeding that of fully supervised approaches. 
DeActive \cite{hossain2018deactive} focuses on reducing the computational and resource overhead of AL-based HAR systems, making them more suitable for deployment on resource-constrained devices.

\textbf{Inexact Supervised Learning:}
Inexact supervised learning arises when the available labels are coarse or ambiguous.
For Inexactly supervised HAR, Siamese Networks \cite{koch2015siamese} provide an effective architecture for generalization.
As shown in Fig \ref{fig:sia}, a pair of samples is fed into two networks that share weights and output correlation discrimination, such as binary discrimination or similarity distance.
KATN \cite{you2021katn} designs a siamese key activity attention network (SAN) to extract semantic and temporal features of IMU data and simultaneously perform key activity spotting using imprecise labels.
Sheng et al. \cite{sheng2020weakly} utilize siamese networks and temporal CNN for simultaneous HAR and activity-based person recognition, trained only on the similarity information of activities and persons without requiring explicit labels.

\subsubsection{\textbf{Unsupervised Learning}}

Unsupervised learning seeks to uncover latent structures and regularities in data without access to any explicit label information. 
From the perspective of generalizable HAR, such methods aim to exploit clustering and association analysis to learn representations that remain robust across users, devices, and environments, thereby enabling recognition in fully unlabeled and heterogeneous scenarios.

\textbf{Clustering Analysis:}
Clustering-based methods partition unlabeled sensor data into groups according to similarity measures under the inductive assumption that samples in the same cluster correspond to the same activity. 
SWL-Adapt \cite{cross_user_aaai23} enhances intra-class similarity and inter-class separability by learning sample weights during domain alignment and employing pseudo-label–driven meta-optimization. 
Ma et al. \cite{ma2021unsupervised} apply the K-means algorithm to generate pseudo-labels, integrating feature representation learning, clustering, and HAR tasks into a unified multi-task optimization framework. 
Rey et al. \cite{rey2017label} construct similarity graphs between new and existing data sources to support unsupervised clustering, thereby enabling HAR systems to adapt dynamically to new environments. 
DWLR \cite{li2024dwlr} achieves cross-domain alignment by reweighting target-domain sample distributions and enforcing intra-class compactness alongside inter-class separation.
UDA-HAR \cite{cross_location_har_2020} explores unsupervised domain adaptation strategies, aligning target-domain samples with the class structure of the source domain through feature alignment, adversarial domain confusion, and pseudo-labeling, thereby improving intra-class consistency and inter-class separability.

\textbf{Association Analysis:}
By contrast, association analysis emphasizes discovery of latent correlations and temporal patterns rather than explicit partitioning. It mines co-occurrence patterns or temporal dependencies that recur across unlabeled streams.
Xia et al. \cite{ubicomp20_har_factory} identified period and action motifs within the initial segments of sensor data and subsequently tracked them using particle filters, enabling robust recognition of repetitive activity patterns. 
In industrial HAR scenarios, Xia et al. \cite{ubicomp19_har_factory} further leveraged factory process instructions as external knowledge to discover hidden associations between sensor streams and activity patterns. 

\begin{figure}[t]
    \centering
    \includegraphics[width=0.9\textwidth]{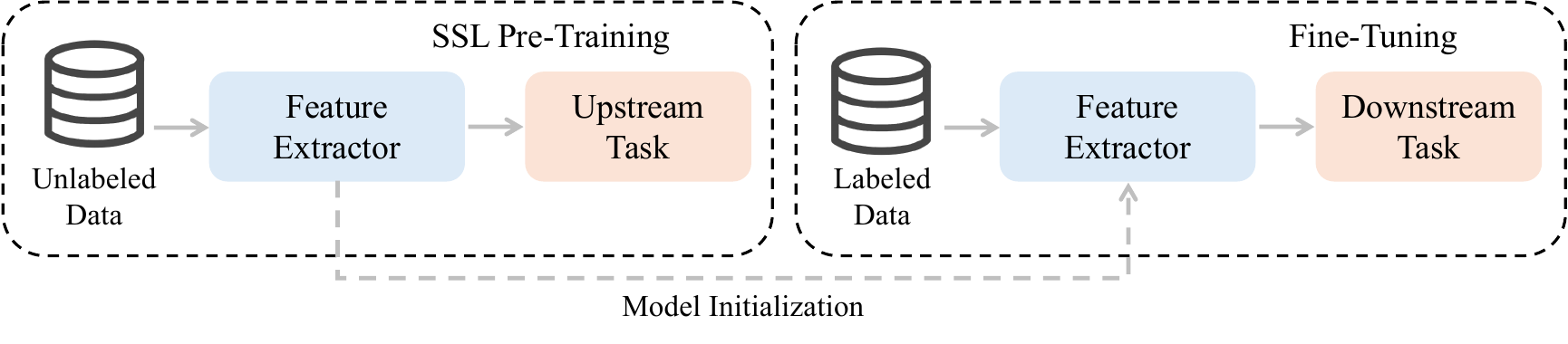}
    \vspace{-0.2in}
    \caption{Illustration of self-supervised learning (SSL) framework.}
    \label{fig:ssl}
\end{figure}

\begin{figure}[t]
    \centering
    \subfloat[Transformation Recognition]{
        \centering
        \includegraphics[width=1.3in]{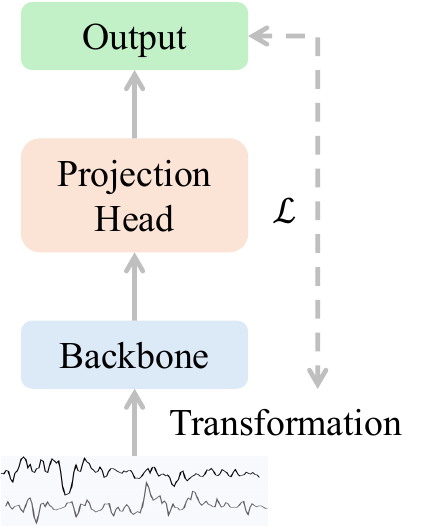}
        \label{fig:self1}
    }
    \hspace{1cm}
    \subfloat[Reconstruction]{
        \centering
        \includegraphics[width=1.in]{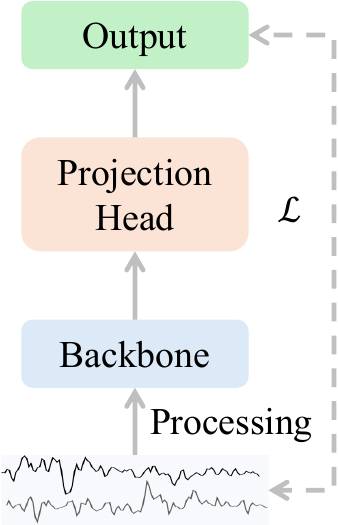}
        \label{fig:self2}
    }
    \hspace{1cm}
    \subfloat[Contrastive Learning]{
        \centering
        \includegraphics[width=1.7in]{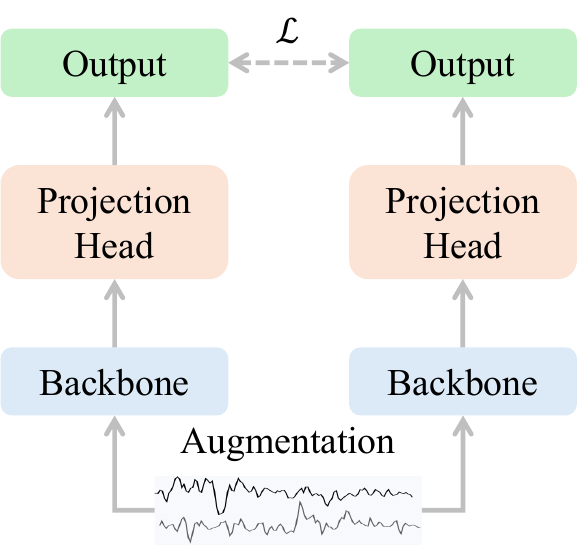}
        \label{fig:self3}
    }
    \caption{Illustration of transformation recognition, reconstruction, and contrastive learning.}
    \label{fig:model1}
    \vspace{-0.1in}
\end{figure}

\subsubsection{\textbf{Self-Supervised Learning}}

Self-supervised learning (SSL) leverages upstream tasks defined on raw sensor streams to learn rich, transferrable representations without relying on manual annotations. 
In IMU-based HAR, SSL is particularly attractive for improving generalization because upstream objectives can 
(i) force models to encode invariances to nuisance factors, 
and (ii) produce embeddings that are easily adapted to downstream tasks with only a small number of labeled data.
A typical SSL pipeline for HAR is illustrated in Fig.~\ref{fig:ssl}.
The mainstream methods include transformation recognition, reconstruction, contrastive learning, and hybrid approaches.

\textbf{Transformation Recognition:}
In transformation recognition–based SSL, the upstream task involves identifying which type of transformation has been applied to an input sensor sequence. 
As illustrated in Fig. \ref{fig:self1}, one or more predefined transformations are first applied to the raw IMU signal. 
The augmented sequence is then fed into a backbone network to extract latent representations, which are subsequently passed through a projection head to produce the final output. 
The objective is to classify the specific transformation applied or to predict its intensity. 
By training on such surrogate objectives, the model learns semantically meaningful and transformation-aware representations that can be transferred to downstream HAR tasks with improved generalization.

Several representative methods have adopted the transformation recognition paradigm to improve generalization in IMU-based HAR. 
Islam et al. \cite{islam2021self} and Tang et al. \cite{tang2021selfhar} apply eight signal transformations to the raw sensor data and train backbones to classify the type of transformation applied.
Temporal Neighborhood Coding (TNC) \cite{tonekaboni2021unsupervised} extends this idea by defining a neighborhood distribution for each signal. 
Given an anchor sequence and a randomly transformed sample, the model predicts whether the augmented signal lies within the anchor's original neighborhood.
MultiTaskSSL \cite{saeed2019multi, haresamudram2022assessing, yuan2024self} recognizes multiple transformations jointly.

\textbf{Reconstruction:}
Inspired by BERT-style pretraining \cite{devlin2019bert}, reconstruction has been adopted in IMU-based HAR to learn robust representations by reconstructing processed portions of the input signal. 
As shown in Fig. \ref{fig:self2}, parts of the raw sensor sequence are processed and then passed through a backbone network to produce latent representations. 
These representations are subsequently decoded via a projection head to reconstruct the signal segments. 
The objective is to minimize the difference between the reconstructed and original signals, encouraging the model to capture the underlying structure and temporal dependencies in the data.

Several reconstruction-based SSL methods have been adapted for IMU-based HAR. 
LIMU-BERT \cite{Xu2021} introduces reading normalization and feature dimension expansion to better align reconstruction with IMU data characteristics. 
Google \cite{narayanswamy2024scaling, xu2025lsm} applies masked reconstruction at scale to develop Wearable Foundation Models.
CRT \cite{zhang2023self} combines time and frequency domains by masking signal segments in both spaces and reconstructing them using a Transformer, capturing cross-domain dependencies. 
STMAE \cite{miao2024spatial} employs an asymmetric encoder-decoder and a two-stage spatiotemporal masking strategy to exploit multi-device correlations and improve performance.

\textbf{Contrastive Learning:}
Contrastive and predictive learning approaches both adopt a multi-branch architecture. 
As shown in Fig. \ref{fig:self3}, two differently augmented views of the same input sample are independently processed by a shared backbone and a projection head to generate latent embeddings. 
The objective is to encourage consistency between these embeddings, enabling the model to learn representations that are invariant to input transformations.

Early contrastive learning methods include Contrastive Predictive Coding (CPC) \cite{haresamudram2021contrastive}, which uses past signal segments as input. 
The left branch of CPC model encodes these segments into latent representations, while the right branch predicts the embeddings of future segments. 
CPC has been extended to various network architectures \cite{haresamudram2023investigating} and combined with techniques such as vector
quantization~\cite{spathis2020learning} to enhance representation quality and stability.
TS-TCC \cite{qian2022makes} combines strong and weak augmentations of the input and uses shared encoders to produce embeddings. 
It predicts future weak embeddings from past strong ones.

Qian et al. apply SimCLR \cite{qian2022makes} to IMU-based HAR by generating two augmented views of each signal and encoding them with a shared backbone and projection head. 
Positive pairs are formed from the same signal, and negative pairs from different signals, with the objective of maximizing agreement within positives while separating negatives.
colloSSL \cite{jain2022collossl} extends this framework to the device level by sampling positive and negative devices, enabling representation learning from unlabeled multi-device data.
CrossDomainHAR~\citep{cross_user_position_TIST25} combines SimCLR with self-training process to extract additional knowledge specific to the target domain.
SSCL \cite{saeed2020federated} applies contrastive learning based on wavelet transform to federated learning.
    
\textbf{Hybrid Approaches:}
Hybrid approaches leverage the combination of multiple upstream tasks or the integration of both upstream and downstream tasks to learn generalizable representations.
CrossHAR \cite{crosshar} and ARSSL \cite{xu2023augmentation} integrate contrastive learning and reconstruction via hierarchical training.
SelfHAR \cite{tang2021selfhar} jointly optimizes multiple transformation recognition tasks with HAR to enhance feature quality.

\subsubsection{\textbf{LLM-based Generalization Learning}}

Recent studies have explored utilizing large language models (LLMs) for enhancing the generalization capability of IMU-based HAR, considering their advanced semantic understanding, reasoning capabilities, and cross-domain transferability.
By integrating an IMU feature encoder with an LLM, the raw time-series signals can be mapped to semantic representations that align with natural language descriptions of activities. 
This enables the model to exploit the rich prior knowledge of human motion patterns acquired from large-scale text corpora. 
Consequently, it becomes possible to perform zero-shot and few-shot recognition of previously unseen activities, as well as to generate diverse IMU data through the LLM.

IMUGPT \cite{leng2023generating} and IMUGPT 2.0 \cite{leng2024imugpt} explore LLM-based data generation by leveraging ChatGPT’s natural language generation capabilities to automatically produce textual descriptions of activities. 
These descriptions are then combined with motion synthesis and signal processing techniques to generate virtual IMU data streams. 
HARGPT \cite{ji2024hargpt} investigates the use of simple prompts to directly input raw sensor data into an LLM, enabling exploration and understanding of the model’s capabilities for HAR. 
This approach achieves excellent HAR performance and demonstrates strong robustness. 
UbiPhysio \cite{wang2024ubiphysio} focuses on fine-grained action descriptions and utilizes the commonsense knowledge of LLMs to generate professional feedback. 
Sensor2Text \cite{chen2024sensor2text} exploits the powerful conversational abilities of LLMs to perform question–answering tasks across various wearable sensor modalities.

\subsection{Data-Centric Methodology}
\label{sec:data}

To enhance the performance and generalizability, recent research has enhanced IMU-based HAR models from a data-centric perspective, including strategies such as multi-modal fusion, cross-modal learning, and data augmentation.
In multi-modal fusion and cross-modal learning approaches, IMU data can be fused with other modalities to leverage complementary information, or alternative modalities can be utilized to strengthen IMU-based models and improve their generalization.
In addition, data augmentation techniques have been widely adopted to expand the training set and introduce variability, helping models better handle unseen conditions and reduce overfitting.

\subsubsection{\textbf{Multi-Modal Fusion}}

Multi-modal fusion methods combine IMU data with other sensor modalities to enhance the generalization of HAR models through joint representation learning. 
A common architecture involves extracting features from each modality using dedicated encoders, followed by a fusion network that integrates the modality-specific representations into a unified embedding for HAR.

Many works have leveraged multi-modal fusion to improve HAR performance by integrating IMU data with other sensor modalities. 
MESEN \cite{xu2023mesen} introduces a multi-modal single-modality-aware framework, where fusion during pretraining helps optimize unimodal feature extractors. 
EgoLocate \cite{yi2023egolocate} combines six IMUs with a monocular smartphone camera for real-time motion capture. 
EgoADL \cite{sun2024multimodal} fuses non-visual signals, including audio, wireless sensing, and motion sensors to support HAR. 
HabitSense \cite{fernandes2024habitsense} integrates RGB, thermal, and IMU data for health HAR. 
MMTSA \cite{gao2023mmtsa} transforms IMU signals into grayscale images and fuses them with visual data through an inter-segment attention module. 
iMove \cite{liu2024imove} demonstrates the benefit of combining bio-impedance with IMU data, while Bhattacharya et al. \cite{bhattacharya2022leveraging} and SAMoSA \cite{mollyn2022samosa} incorporate voice signals to enhance HAR.
Cosmo \cite{ouyang2022cosmo} introduces a quality-guided attention mechanism to extract both consistent and complementary information across IMU and video modalities, enabling more effective fusion.

In addition, some studies treat IMU sub-modalities, such as accelerometer, gyroscope, and magnetometer data, as distinct input streams. 
IF-ConvTransformer \cite{zhang2022if} fuses these signals based on their physical relationships. 
DeepSense \cite{Yao2017} combines CNN and RNN modules for IMU sub-modality fusion, while AttnSense \cite{Ma2019} employs attention mechanisms with CNN and GRU layers. 
DynamicWHAR \cite{miao2022towards} further models dynamic interactions across IMU sensor modalities to adaptively aggregate inter-sensor information.

\subsubsection{\textbf{Cross-modal Learning}}

Given the high cost and labor intensity associated with annotating IMU data, cross-modal learning has emerged as a promising strategy to enhance IMU-based HAR. 
This approach leverages richly labeled datasets or model capabilities from other modalities to learn transferable representations that benefit modalities with limited labeled data, particularly IMU.

Recent approaches explore cross-modal translation between IMU data and other modalities. 
Sensor2Text \cite{chen2024sensor2text} leverages visual-language models (VLMs) and LLMs to jointly process sensor and textual data for HAR and interactive question answering (QA).
Vision2Sensor \cite{radu2019vision2sensor} transfers knowledge from visual modalities to on-device sensors. 
IMU2Doppler \cite{bhalla2021imu2doppler} uses labeled IMU data to train a mmWave radar-based HAR system with minimal supervision.
IMUGPT \cite{leng2023generating} and IMUGPT2.0 \cite{leng2024imugpt} generate synthetic motion signals guided by LLMs.
HARGPT \cite{ji2024hargpt} uses raw IMU as LLM input, employing role-playing and chain-of-thought prompting for HAR.
ZARA \cite{li2025zara} is an agent-based framework designed for zero-shot, interpretable HAR directly from raw motion time series.
UbiPhysio \cite{wang2024ubiphysio} transforms IMU activity data into descriptive text for expert feedback. 
DeXAR \cite{arrotta2022dexar} converts sensor signals into semantic images to enhance interpretability through explainable AI techniques.
X-CHAR \cite{jeyakumar2023x} models complex activities as human-interpretable concept sequences. 
Generative approaches like IMUTube \cite{Kwon2020}, Vi2IMU \cite{santhalingam2023synthetic}, and the work of Kwon et al. \cite{kwon2021approaching} synthesize IMU signals from video.

Meanwhile, there are other works that focus on cross-modal contrastive. 
Tong et al. \cite{tongzeroshot21} propose a video-based semantic space to associate seen and unseen activity categories via video embeddings. 
IMU2CLIP  \cite{moon2023imu2clip} and PRIMUS \cite{das2024primus} align sensor features with video and text representations. 
GOAT \cite{cross_dataset_miao2024goat} leverages textual attributes from activity labels and sensor locations for wearable data pretraining. 
SensorLM \cite{zhang2025sensorlm} introduces a hierarchical subtitle generation scheme to better align motion and language. 
Similarly, SensorLLM \cite{li2024sensorllm} aligns sensor data with trend descriptions.
VAX  \cite{patidar2023vax}  uses multimodal (video/audio) supervision to annotate motion data. 
COMODO \cite{chen2025comodo} leverages a video encoder to construct a dynamic instance queue, thereby aligning the feature distributions of video and IMU embeddings.
TS2ACT \cite{xia2024ts2act} aligns time series and images using contrastive learning between a pretrained CLIP image encoder and a temporal encoder.
FOCAL \cite{liu2023focal} imposes a temporal structural constraint on modality features, leveraging the mean distance between samples with large temporal intervals to constrain that between temporally adjacent samples for alignment.
COCOA \cite{deldari2022cocoa} employs a novel objective function that learns high-quality representations by computing the cross-correlation between different data modalities and minimizing the similarity between uncorrelated instances.
CroSSL \cite{deldari2024crossl} leverages improved cross-modal contrastive learning, aiming to support handling of missing modalities and end-to-end cross-modal learning.

\subsubsection{\textbf{Data Augmentation}} 

Data scarcity and limited diversity remain major challenges to the generalizability of HAR models. 
To address these limitations, many studies have explored data augmentation techniques tailored to IMU-based sensing. 

Methods such as CrossHAR \cite{crosshar}, UniHAR \cite{unihar}, and UniMTS \cite{zhang2024unimts} investigate sensor data generation principles to enrich the data distribution and improve model robustness. 
AutoAugHAR \cite{zhou2024autoaughar} considers the unique characteristics of candidate augmentation operations as well as the specific challenges of HAR tasks, and proposes a gradient-based two-stage augmentation optimization framework. 
ALAE-TAE-CutMix+ \cite{ahmad2023alae} first enhances latent information within each sensor channel and introduces a novel augmentation strategy to address limitations in channel-wise transformations. 
Abedin et al. \cite{abedin2021attend} examine task-agnostic augmentations for sensor streams as a regularization strategy. 
Oppel et al. \cite{oppel2025diffusion} leverage denoising diffusion probabilistic models to synthesize IMU data generalizable to unseen subjects. 

rTsfNet \cite{enokibori2024rtsfnet} performs feature-level augmentation via automatic extraction of multiple 3D bases from rotational parameters. 
ConvBoost \cite{shao2023convboost} introduces three augmenters, R-Frame (dense resampling), Mix-up (synthetic interpolation), and C-Drop (channel-wise dropout), to enhance sample diversity across training epochs. 
HMGAN \cite{chen2023hmgan} uses generative adversarial networks to maintain both modality-specific details and global consistency. 
SHARE \cite{zhang2023unleashing} introduces three label-space augmentation strategies: a basic token-level method, and embedding-level and sequence-level augmentations guided by pretrained models. 
Zhou et al. \cite{zhou2025learning} model the bias in IMU readings as a probability distribution and employed a conditional diffusion model to approximate this distribution. 
SeRaNDiP \cite{Kalupahana2023SeRaNDiP} exploits the inherent stochastic noise in wearable sensors for privacy-preserving augmentation that also improves generalization.

\subsection{Datasets and Benchmarks}
\label{sec:datasets}

Benchmark datasets play a critical role in advancing IMU-based HAR by providing standardized data for fair algorithm evaluation, enabling reproducible experiments, and fostering comparative studies across methods.
In this section, we introduce 25 widely used datasets, summarized in Table \ref{tab:datasets}, along with key attributes such as the number of subjects, sensor types, activity categories, data volume, and publication year. 
To facilitate comparison and highlight differences in data characteristics, we categorize these datasets into two major groups: (i) IMU-only datasets and (ii) multimodal datasets, where the latter typically combine IMUs with other sensor modalities such as vision, audio, or physiological signals.
In addition, we organize some IMU-based HAR-related tools and benchmarks to assist research in this field.

\subsubsection{\textbf{IMU-Only Datasets}}

IMU-only datasets focus exclusively on inertial signals, such as accelerometer, gyroscope, and sometimes magnetometer data, collected from wearable sensors.
Despite the absence of complementary modalities, many IMU-only datasets cover a diverse range of activities and sensor configurations, making them valuable for developing and benchmarking generalizable HAR models.

\begin{itemize}

    \item \textbf{CAPTURE-24~\cite{chan2024capture}:}  
    The CAPTURE-24 dataset comprises wrist-worn accelerometer data collected from 151 participants over approximately 24-hour periods. 
    To obtain reliable ground truth annotations, participants were equipped with both a wrist-mounted accelerometer and a body-worn camera during waking hours. 
    Additionally, sleep diaries were used to document individual sleep periods.
    The accelerometer used was the Axivity AX3, a tri-axial device worn on the wrist, sampling acceleration data at 100 Hz. Ground truth activity labels were derived from images captured by the Vicon Autographer, a wearable camera that recorded one image every 20 seconds. 
    It is important to note that the dataset only provides the raw accelerometer signals along with time-aligned textual annotations derived from manual image reviews.

    \item \textbf{TNDA-HAR~\cite{yan2021tndahar}:} 
    The TNDA-HAR dataset consists of inertial sensor data, including tri-axial accelerometer and gyroscope measurements, collected from 23 participants engaged in eight common daily activities such as walking, running, cycling, sitting, standing, stair ascent/descent, and lying. 
    Data were captured in both free-living and laboratory environments, with sensors placed at multiple body locations, including the wrist, ankle, and back.
    Each participant contributed approximately 14.6 minutes of continuous activity recordings, resulting in a total dataset duration of around 5.7 hours. 
    Designed to facilitate research in HAR, the dataset is particularly suited for exploring topological and nonlinear dynamical analysis methods.

    \item \textbf{HAR70+~\cite{ustad2023validation}:} 
    The HAR70+ dataset comprises inertial recordings from 18 older adults, aged between 70 and 95 years, captured in a semi-structured free-living environment. 
    Each participant wore two Axivity AX3 accelerometers, each providing tri-axial acceleration data at a sampling rate of 50 Hz, for approximately 40 minutes. 
    One sensor was positioned on the right front thigh and the other on the lower back to capture lower-body movement dynamics. 
    Notably, five participants utilized walking aids during data collection.
    Ground truth activity annotations were obtained through frame-by-frame analysis of video footage recorded via a chest-mounted camera.

    \item \textbf{WISDM~\cite{weiss2019wisdm}:} 
    The WISDM Smartphone and Smartwatch Activity and Biometrics Dataset comprises tri-axial accelerometer and gyroscope data collected from 51 participants who concurrently wore a smartphone and a wrist-mounted smartwatch. 
    Each participant performed 18 common daily activities, including walking, jogging, stair ascent/descent, brushing teeth, and folding clothes, with each activity recorded for approximately three minutes. 
    The dataset is designed to support sensor-based HAR by leveraging data from both smartphone and wearable devices.

    \item \textbf{MotionSense~\cite{malekzadeh2018protecting}:} 
    The MotionSense dataset consists of time-series data captured from the accelerometer and gyroscope sensors of an iPhone 6s, sampled at 50 Hz. 
    The smartphone was placed in the front pocket of each participant throughout data collection.
    A total of 24 participants, varying in gender, age, weight, and height, performed six distinct activities—walking, jogging, stair ascent, stair descent, sitting, and standing—across 15 repeated trials, all conducted under consistent environmental conditions. 
    In addition to facilitating traditional HAR tasks, the dataset is also intended to support research on identifying biometric and behavioral patterns embedded in sensor time-series data, such as inferring individual attributes like gender or personality traits.

    \item \textbf{SHL Challenge~\cite{gjoreski2018benchmarking}:} 
    The Sussex-Huawei Locomotion (SHL) Challenge dataset was developed to support the recognition of eight locomotion and transportation modes based on smartphone inertial sensor data. 
    The dataset includes a total of 271 hours of training data and 95 hours of test data, all collected from a single individual using a single smartphone. 

    \item \textbf{MobiAct~\cite{vavoulas2016mobiact}:} 
    The MobiAct dataset comprises inertial data collected via smartphones from 66 participants performing over 3,200 trials. 
    It includes four types of simulated falls, twelve activities of daily living (ADLs), and one extended daily living scenario. 
    The selection of ADLs was guided by three main criteria: 
    (i) fall-like activities in which the subject typically ends in a motionless state, such as sitting down or entering/exiting a vehicle; 
    (ii) sudden or high-dynamic activities that may resemble falls, including jumping and jogging;
    and (iii) common everyday movements such as walking, standing, and stair ascent/descent.

    \item \textbf{Complex Human Activities~\cite{shoaib2016complex}:} 
    The Complex Human Activities dataset comprises tri-axial accelerometer, gyroscope, and magnetometer data collected from 10 participants performing 13 complex daily activities. 
    Each participant carried a smartphone in their trouser pocket and wore a motion sensor on the wrist, enabling multi-position sensing. 
    The recorded activities include walking, jogging, sitting, standing, biking, stair ascent and descent, typing, drinking coffee, eating, giving a talk, and smoking.
    Each activity was performed for 30 minutes per participant, resulting in a total of 390 minutes of annotated sensor data. 
    All signals were sampled at 50 Hz. 

    \item \textbf{HHAR~\cite{stisen2015smart}:} 
    The HHAR dataset comprises two sub-datasets designed to investigate the impact of sensor heterogeneity on HAR tasks. 
    Data were collected using smartwatches and smartphones while participants performed a series of scripted activities in an unspecified order.
    The dataset includes readings from two commonly embedded motion sensors, an accelerometer and a gyroscope, sampled at the maximum frequency supported by each device. 

    \item \textbf{WHARF~\cite{bruno2013analysis}:} 
    The WHARF dataset consists of labeled accelerometer recordings collected using a single wrist-worn tri-axial accelerometer. 
    It is specifically designed to support the development and validation of acceleration-based models for recognizing eight distinct motion primitives. 

    \item \textbf{DSADS~\cite{rajasegarar2013daily}:} 
    The Daily and Sports Activities Dataset (DSADS) comprises multimodal motion sensor data collected from eight participants performing 19 distinct daily and sport-related activities. 
    Each activity was performed for approximately five minutes in a controlled setting. 
    The dataset utilizes five Xsens MTx IMUs, each integrating tri-axial accelerometer, gyroscope, and magnetometer sensors. 
    The sensors were strategically placed on the torso, both arms, and both legs to capture full-body motion dynamics. 
    All signals were recorded at a sampling frequency of 25 Hz.

    \item \textbf{UCI-HAR~\cite{anguita2013public}:} 
    The UCI-HAR dataset comprises inertial sensor data collected from 30 participants performing six predefined activities: walking, walking upstairs, walking downstairs, sitting, standing, and lying. 
    The data were recorded using a smartphone equipped with embedded accelerometer and gyroscope sensors, which the subjects carried during task execution. This dataset has become a widely used benchmark for evaluating HAR algorithms under controlled conditions.

    \item \textbf{USC-HAD~\cite{mi12:ubicomp-sagaware}:} 
    The USC Human Activity Dataset (USC-HAD) was developed to support research in HAR within the ubiquitous computing community. 
    The dataset comprises data collected from 14 participants performing 12 common daily activities. 
    A wearable sensing device, equipped with motion sensors, was attached to the front right hip of each subject during data collection, allowing for consistent and unobtrusive monitoring of body motion.

    \item \textbf{Daphnet Freezing of Gait \cite{bachlin2009wearable}:}
    The Daphnet Freezing of Gait dataset was developed to facilitate the evaluation of algorithms aimed at detecting freezing of gait (FoG) episodes in Parkinson’s disease patients. 
    Data were collected in a laboratory environment using wearable tri-axial accelerometers placed on the legs and hip. 
    The experimental design was intentionally structured to provoke frequent FoG events.
    Participants engaged in three types of tasks: (1) straight-line walking, (2) walking with frequent turns, and (3) a semi-naturalistic activity of daily living scenario that involved tasks such as walking through different rooms, fetching coffee, and opening doors.

    \item \textbf{Skoda Mini Checkpoint \cite{zappi2008activity}:}
    The Skoda Mini Checkpoint dataset captures sensor data from assembly-line workers performing quality control tasks in an automotive manufacturing environment. 
    A single worker was instrumented with 19 accelerometers placed on both arms while executing 10 distinct factory activities. 
    The dataset includes approximately 3 hours of recordings at a sampling rate of 98 Hz, with each gesture repeated around 70 times. 

\end{itemize}

IMU-only datasets provide a foundational testbed for developing efficient HAR models with potential for real-world deployment. 
However, their limited sensor diversity and often constrained recording environments may pose challenges for generalization across users, devices, and contexts. 
Achieving robust generalization with IMU-only data often requires careful design of cross-domain evaluation methods, as well as data augmentation or domain adaptation strategies.

\begin{table*}[t]
    \caption{Summary of HAR dataset.}
    \label{tab:datasets}
    \centering
    \small
    \resizebox{0.95\textwidth}{!}{
    \begin{tabular}
    {m{3cm}<{\centering}  m{1cm}<{\centering} m{4cm}<{\centering} m{4cm}<{\centering} m{1cm}<{\centering} m{1cm}<{\centering}}
    \toprule
    \textbf{Dataset} &
    \textbf{Subjects} &
    \textbf{Sensors} &
    \textbf{Activities} &
    \textbf{datasize} &
    \textbf{Year} 
    \\
    \midrule
    \multicolumn{6}{c}{\textbf{IMU-Only Datasets}}  \\
    \midrule
    
    CAPTURE-24~\cite{chan2024capture} &
    151 &
    acc &
    200 unique labels &
    3883 h  &
    2024  \\ 
    \midrule

    TNDA-HAR~\cite{yan2021tndahar} &
    23 &
    acc, gyro &
    8 daily activities &
    5.7 h &
    2021  \\
    \midrule

    HAR70+ ~\cite{ustad2023validation} &
    18 &
    acc &
    8 daily activities &
    12.6 h &
    2020  \\
    \midrule
    
    WISDM~\cite{weiss2019wisdm} &
    51 &
    acc, gyro &
    18 daily activities &
    91.8 h &
    2019  \\ 
    \midrule   

    MotionSense~\cite{malekzadeh2018protecting} &
    24 &
    acc, gyro &
    6 daily activities &
    - &
    2019  \\ 
    \midrule

    SHL Challenge~\cite{gjoreski2018benchmarking} &
    3 &
    acc, gyro, mag & 
    8 transportation modes &
    2812 h &
    2018  \\ 
    \midrule

    MobiAct~\cite{vavoulas2016mobiact} &
    57 &
    acc, gyro & 
    9 daily activities and 4 falls & 
    - &
    2016  \\ 
    \midrule

    Shoaib~\cite{shoaib2016complex} &
    10 &
    acc, gyro, mag &
    13 daily activities &
    6.5 h &
    2016  \\
    \midrule

    HHAR~\cite{stisen2015smart} &
    9 &
    acc, gyro & 
    6 daily activities &
    - &
    2015  \\ 
    \midrule

    WHARF~\cite{bruno2013analysis} &
    16 &
    acc &
    8 motion primitives &
    - &
    2013  \\
    \midrule

    DSADS~\cite{rajasegarar2013daily} &
    8 &
    acc, gyro, mag & 
    19 daily, sports activities &
    12.7 h  & 
    2013  \\
    \midrule

    UCI-HAR~\cite{anguita2013public} &
    30 &
    acc, gyro &
    6 daily activities &
    - &
    2012  \\ 
    \midrule

    USC-HAD~\cite{mi12:ubicomp-sagaware} &
    14 &
    acc, gyro, mag &
    12 daily activities &
    - &
    2012  \\
    \midrule

    Daphnet FoG~\cite{bachlin2009wearable} &
    10 &
    acc &
    3 walking activities &
    8.3 h &
    2009  \\
    \midrule

    Skoda Mini Checkpoint~\cite{zappi2008activity} &
    1 &
    acc, 3D acc sensor &
    10 assembly-line activities &
    - &
    2008  \\
    
    \midrule
    \multicolumn{6}{c}{\textbf{Multimodal Datasets}}  \\
    \midrule

    WEAR~\cite{bock2024wear} &
    22 &
    acc, video & 
    18 sports activities &
    19 h &
    2024  \\ 
    \midrule

    HARTH~\cite{logacjov2021harth} &
    22 &
    acc, video &
    12 daily activities &
    35.9 h &
    2021  \\
    \midrule

    w-HAR~\cite{bhat2020whar} &
    22 &
    acc, gyro, stretch sensor &
    7 daily activities &
    3 h & 
    2020  \\
    \midrule

    MMAct~\cite{Kong_2019_ICCV} &
    40 &
    RGB-video, keypoints, acc, gyro, ori, Wi-Fi, pressure &
    37 daily, abnormal, desk work activities &
    - & 
    2019  \\
    \midrule

    RealWorld HAR~\cite{sztyler2016body} &
    15 &
    acc, gyro, mag, GPS, light, sound level & 
    8 daily activities &
    124.3h &
    2016  \\ 
    \midrule

    UTD-MHAD~\cite{chen2015utd-mhad} &
    8 &
    RGB video, depth video, skeleton positions, acc, gyro &
    27 daily, sports activities, gestures &
    - & 
    2015  \\
    \midrule

    MHEALTH~\cite{banos2014mhealth} &
    10 &
    acc, gyro, mag, ECG &
    12 daily activities &
    - & 
    2014  \\
    \midrule

    Berkeley MHAD~\cite{Berkeley-MHAD} &
    12 &
    acc, optical capture system, video, depth sensor, audio &
    11 daily activities &
    1.37 h &
    2013  \\
    \midrule
    
    PAMAP2~\cite{reiss2012introducing} &
    9 &
    acc, gyro, mag, heart rate &
    18 daily activities &
    10 h &
    2012  \\ 
    \midrule
    
    Opportunity~\cite{roggen2010collecting} &
    4 &
    acc, gyro, mag, ambient sensors &
    9 kitchen activities, 9 gestures &
    25 h &
    2011  \\ 
    \bottomrule 
    
    \end{tabular}
    }
\end{table*}

\subsubsection{\textbf{Multimodal Datasets}}

Multimodal datasets incorporate IMU signals alongside additional sensing modalities such as RGB video, depth data, audio, or physiological signals. 
These datasets aim to capture richer contextual and semantic information about human activities, enabling more accurate and robust HAR models. 

\begin{itemize}

    \item \textbf{WEAR~\cite{bock2024wear}:} 
    The Wearable Egocentric Activity Recognition (WEAR) dataset is a multi-modal dataset designed for HAR in outdoor sports contexts. 
    It comprises synchronized inertial and egocentric video data collected from 22 participants performing 18 distinct workout activities across 11 outdoor locations. 
    The inertial data include tri-axial acceleration signals, while the visual data consist of first-person video recordings. 

    \item \textbf{HARTH~\cite{logacjov2021harth}:} 
    The Human Activity Recognition Trondheim (HARTH) dataset is a professionally annotated dataset designed to support research in free-living HAR. 
    It includes data from 22 participants, each wearing two tri-axial accelerometers, one mounted on the right thigh and the other on the lower back, for approximately two hours. 
    Data were collected in naturalistic, uncontrolled environments to better reflect real-world behavior.

    \item \textbf{w-HAR~\cite{bhat2020whar}:} 
    The w-HAR dataset integrates tri-axial accelerometer and gyroscope data sampled at 100 Hz with wearable stretch-sensor measurements to support multimodal HAR. 
    Data were collected from 22 participants performing seven common activities: jumping, lying down, sitting, standing, stair ascent, stair descent, and walking. 
    The raw sensor streams were segmented into 4,740 single-activity windows, amounting to approximately three hours of annotated data.
    The dataset is publicly available alongside a complete HAR framework covering segmentation, feature extraction, classifier design, and online adaptation methods.

    \item \textbf{MMAct~\cite{Kong_2019_ICCV}:} 
    The MMAct dataset is a large-scale, multi-modal dataset for HAR, comprising over 36,000 trimmed video clips spanning 37 daily action classes performed by 40 participants. 
    Data were collected across seven sensing modalities: RGB videos, human keypoints, accelerometer, gyroscope, orientation, Wi-Fi signals, and pressure data. 
    Recordings were conducted in diverse scenes. 

    \item \textbf{RealWorld HAR~\cite{sztyler2016body}:} 
    The RealWorld HAR dataset provides comprehensive multimodal sensor data collected from 15 participants performing a variety of common daily activities. 
    The recorded modalities include tri-axial acceleration, GPS, gyroscope, ambient light, magnetic field, and sound level. 
    To ensure full-body motion coverage, sensors were simultaneously placed on seven body locations: chest, forearm, head, shin, thigh, upper arm, and waist.
    Participants engaged in eight activity classes: walking, running/jogging, sitting, standing, lying, stair ascent, stair descent, and jumping. 
    Most activities were performed for approximately 10 minutes per subject, except for jumping. 
    The dataset is gender-balanced to ensure demographic representativeness.

    \item \textbf{UTD-MHAD~\cite{chen2015utd-mhad}:} 
    The UTD Multimodal Human Action Dataset (UTD-MHAD) contains 861 action sequences representing 27 distinct gestures, performed by eight participants. 
    Each action was repeated four times per subject. 
    The dataset features two synchronized modalities: (i) RGB-D and skeleton data acquired via Microsoft Kinect V2, and (ii) inertial data collected from a wearable motion sensor worn on the right wrist.

    \item \textbf{MHEALTH~\cite{banos2014mhealth}:} 
    The MHEALTH dataset contains multimodal recordings of body motion and vital signs collected from ten participants performing twelve physical activities in an out-of-laboratory environment. 
    The activities span a wide range of motion types, including static postures (e.g., standing still, sitting, lying down), dynamic actions (e.g., walking, climbing stairs, cycling, jogging, running, jumping), and limb-specific movements (e.g., frontal arm elevations, waist bends, knee bends).
    Data were acquired using three Shimmer2 wearable sensing units placed on the chest, right wrist, and left ankle. 
    Each unit recorded tri-axial acceleration, gyroscope, and magnetometer data, while the chest sensor also collected a two-lead ECG signal. 
    All signals were sampled at 50 Hz, resulting in approximately 72.1 MB of synchronized time-series data.

    \item \textbf{Berkeley MHAD \cite{Berkeley-MHAD}:}
    The Berkeley Multimodal Human Action Database (MHAD) dataset provides time-synchronized and geometrically calibrated recordings from a diverse set of sensing modalities, including an optical motion capture system, multi-baseline stereo cameras from multiple views, depth sensors, accelerometers and microphones. 
    It contains 11 predefined human actions performed by 12 subjects (7 male and 5 female), predominantly aged between 23 and 30 years, with the inclusion of one elderly participant.
    Each subject completed five repetitions of every action, resulting in approximately 660 multimodal action sequences and a total recording duration of around 82 minutes.

    \item \textbf{PAMAP2~\cite{reiss2012introducing}:} 
    The PAMAP2 dataset comprises recordings of 18 distinct physical activities, such as walking, cycling, and playing soccer, performed by nine participants. 
    Each subject wore three IMUs positioned on the wrist, chest, and ankle, along with a heart rate monitor, capturing synchronized multimodal physiological and motion data.
    This dataset can be used for HAR and intensity estimation.

    \item \textbf{Opportunity~\cite{roggen2010collecting}:}  
    The OPPORTUNITY dataset is designed to benchmark HAR algorithms, encompassing tasks such as classification, automatic data segmentation, sensor fusion, and feature extraction. 
    It integrates three types of sensors: wearable sensors attached to the users, object-embedded sensors within the environment, and ambient sensors capturing contextual information. 
    Activities performed by the users are annotated at multiple hierarchical levels
    
\end{itemize}

Multimodal datasets offer richer contextual cues and complementary information, which can substantially enhance the generalization capacity of HAR models. 
By combining IMU signals with other modalities, these datasets enable more robust feature learning and better disambiguation of similar activities.
However, the increased complexity and modality dependencies can reduce applicability in resource-constrained or wearable-only scenarios. 
Generalizing across modalities, devices, and environments remains a critical research challenge, motivating the need for scalable fusion architectures and modality-agnostic learning approaches.

\subsubsection{\textbf{Tools and Benchmarks}}

To advance IMU-based HAR, some tools and benchmarking frameworks have been proposed to support data preprocessing, model development, and standardized evaluation. 
For instance, exHAR \cite{kianpisheh2024exhar} is a knowledge-driven, interpretable HAR toolkit that allows users to define activities as logical propositions and debug models through explanation and corrective feedback. 
AIP-Net \cite{tanigaki2022predicting} estimates potential performance improvements before additional data collection, aiding efficient retraining decisions. 
CrowdAct \cite{mairittha2021crowdact} introduces a gamified active learning strategy for accurate labeling via crowdsourcing, while SenseCollect \cite{chen2021sensecollect} investigates the influence of data collection settings through controlled experiments.

Furthermore, Hiremath et al. \cite{hiremath2020deriving} propose a complexity assessment framework that quantifies HAR task difficulty via vector space representations. 
The Wearables Development Toolkit \cite{haladjian2019wearables} offers an integrated environment for annotating, analyzing, and evaluating HAR applications on wearable devices. 
Additionally, Li et al. \cite{li2017progress} present a sensor-based system for real-time modeling and progress estimation of structured workflows.
Haresamudram et al. \cite{haresamudram2025past} evaluate sensor-based HAR technologies and provide practical tutorials to guide practitioners in developing HAR systems for real-world application scenarios.

\subsection{Applications of IMU-based HAR}
\label{sec:application}

IMU-based HAR has been widely adopted across various application domains. 
Typical devices used for IMU sensing include smartphones, smartwatches, smart rings, fitness trackers, and dedicated inertial sensor units placed on different body parts (e.g., wristbands, chest straps, ankle sensors). 
The broad applicability of IMU-based HAR can be categorized into several key areas.

\begin{itemize}[leftmargin=*]

    \item \textbf{Healthcare and Rehabilitation:}
    IMU sensors facilitate continuous monitoring of patients' physical activities, enabling precise assessment of rehabilitation progress, detection of abnormal behaviors (e.g., falls), and remote patient monitoring \cite{choi2022deep, inoue2019integrating, kong2013development, wang2024ubiphysio, ley2024co, fernandes2024habitsense, arakawa2023lemurdx, wang2021leveraging}. 
    For instance, Inoue et al. \cite{inoue2019integrating} propose an integrated system for HAR and nursing home care record collection. 
    This system utilizes smartphones to gather activity labels and sensor data, and they conducted a four-month experiment in a nursing home to evaluate its effectiveness.
    
    \item \textbf{Sports and Fitness Monitoring:}
    Athletes and fitness enthusiasts utilize IMU-equipped wearable devices to track their daily activities, exercise intensity, and performance metrics \cite{cauchard2019positive, boovaraghavan2023tao, bock2024wear, khan2017activity}. 
    Khan et al. \cite{khan2017activity} propose an automated cricket batting recognition framework that classifies batting actions based on IMU data to evaluate stroke quality. 
    The system also generates visualizations of critical performance metrics, providing coaches and domain experts with actionable insights to guide technique refinement and performance improvement.

    \item \textbf{Occupational Work Assessment:}
    IMU-based HAR systems can objectively quantify several latent aspects of work, such as task efficiency, spatial utilization, and social interaction patterns \cite{ubicomp19_har_factory,ubicomp20_har_factory, yoshimura2022acceleration, di2020multi, montanari2017detecting}. 
    For instance, Montanari et al. \cite{montanari2017detecting} analyze team dynamics and the strength of employees' connections to space use and organizational hierarchy using data collected through wearable devices at a company adopting activity-based working principles.

    \item \textbf{Smart Home and Assisted Living:}
    HAR systems based on IMU sensors support smart home applications by recognizing residents' daily activities and providing context-aware services, such as adaptive environmental control, daily monitoring, and personalized health \cite{yu2022thumbup, zhang2021fine, hiremath2022bootstrapping}. 
    Hiremath et al. \cite{hiremath2022bootstrapping} propose an approach to constructing HAR systems tailored for individual smart homes. 
    The system passively observes human activities and extracts rich representations from raw sensor data. 
    These representations are then aggregated into activity models using motif learning, requiring only minimal supervision during the learning process.

    \item \textbf{Transportation and Mobility:}
    In transportation and mobility domains, IMU sensors embedded in smartphones or dedicated wearable devices help detect transportation modes, analyze driver behavior, and community-building \cite{gjoreski2018benchmarking, chen2023enhancing, hong2024smallmap, hong2024nationwide, hong2023urban, hong2023autobuild, hwang2024more}. 
    Smallmap \cite{hong2024smallmap} focuses on generating community road networks within enclosed areas such as residential districts. It enhances the process using multimodal sensor data collected during deliveries through a trajectory detection module. Furthermore, by employing a dual spatio-temporal generative adversarial network module, it performs unsupervised road network adaptation and integrates points-of-interest trajectories to automatically generate the road network.

    \item \textbf{Human-Robot Interaction:}
    IMU-based HAR is increasingly applied in human-robot interaction scenarios \cite{cifuentes2014human, gromov2019proximity, belcamino2025distributed}, where recognizing human gestures and activities allows robots to better understand human intentions and respond appropriately. 
    Belcamino et al. \cite{belcamino2025distributed} introduce a HAR system that integrates a modular IMU-equipped data glove with a vision-based tactile sensor to capture detailed hand movements during physical interaction with robots. 
    Similarly, Gromov et al. \cite{gromov2019proximity} propose a system for co-located human–robot interaction that utilizes IMU-sensed pointing gestures to facilitate intuitive communication with mobile robots.
    
    \item \textbf{IMU-based AR and VR Interactions}:
    IMU-based motion capture has made great strides, which can help address VR’s need for affordable, occlusion-free body tracking. 
    Emerging efforts from both academia~\cite{xu2024mobileposer,arakawa2023mi,streli2023hoov,islam2024motiontrace,oh2024uwb} and industry (e.g. Meta Reality~\cite{tsutsui2021you,aria2023dataset}) are beginning to apply lightweight IMU-based tracking 
    to drive user avatars in social VR, control game characters, and support fitness or rehabilitation training in VR without relying on cameras.
    MobilePoser~\cite{xu2024mobileposer} uses the IMUs in everyday mobile devices to enable on-the-go motion tracking. 
    HOOV~\cite{streli2023hoov} proposes an IMU sensing-based spatial interaction technique where the user’s physical space becomes an interaction canvas. 
    Despite these explorations, challenges such as sensor drift, limited accuracy, and long-duration full-body tracking remain unsolved, 
    making IMU-driven AR and VR interaction a promising research direction.
    
    \item \textbf{IMU-based sensing for Embodied Agents:} IMU sensors have also been deployed into embodied agent systems (e.g., robotic arm~\cite{huang2025handcept}, legged robots~\cite{yang2023multi,katayama2025learning}, and autonomous vehicles~\cite{alaba2024gps}). In these systems, IMU signals provide high-frequency, low-latency measurements of orientation and acceleration, and are often fused with other sensing modalities such as cameras, LiDAR, and joint encoders, enabling more accurate state estimation and dynamic control for embodied agents.

\end{itemize}
\section{Challenges and Future Directions}
\label{sec:challenge}

\subsection{Challenges and Open Problems}

Despite the notable progress achieved in recent years, IMU-based HAR continues to encounter several challenges and unresolved problems. 
This section outlines the key obstacles and open research questions that remain to be addressed to advance the field toward more robust and universally applicable solutions.

\subsubsection{\textbf{Reliable Annotation and Dataset Bias}}

For IMU-based HAR,
obtaining large-scale and high-quality labeled training data remains a significant bottleneck. 
Manual annotation of activities is labor-intensive and prone to error. 
Additionally, many datasets rely on scripted or controlled laboratory scenarios, which can not fully capture the full variability and diversity of real life. 
As a result, models trained on them can be biased or overly optimistic in performance.
While semi-supervised and SSL approaches \cite{logacjov2024self} help reduce reliance on labeled data, they do not eliminate the need for labeled examples entirely.
Techniques such as active learning \cite{settles2009active}, which queries users for labels when model uncertainty is high, and weak supervision \cite{dehghani2017neural}, which leverages contextual or semantic signals for labeling, offer promising alternatives but remain underexplored in the context of HAR.
Thus, from a data-centric perspective, two open challenges remain for IMU-based HAR tasks: 
\textit{(i) how to obtain abundant and high-quality labels in the wild}, and \textit{(ii) how to ensure model robustness and mitigate bias when potential label errors exist.}

\subsubsection{\textbf{Efficient Personalization under Domain Shift}} 

Although many recent studies address cross-user and cross-device generalization, these problems remain far from resolved. 
The performance of HAR models often degrades when deployed to user populations that differ substantially from the training set. 
Individual differences in movement patterns necessitate personalization to achieve optimal accuracy.
Few-shot learning \cite{cross_user_position_TIST25}, meta-learning techniques \cite{vilalta2002perspective}, and on-device adaptation \cite{LLM4HAR} are being explored to address the domain shift problems and enhance the personalization. 
Test-time adaptation is also a related concept, involving real-time model adjustment during inference using unlabeled input data. 
Wang et al. \cite{wang2024optimization} proposed an optimization-free test-time adaptation method for cross-person HAR that adjusts normalization layers, achieving improved accuracy. 
Despite these efforts, the major challenge lies in \textit{achieving efficient personalization under domain shifts in IMU-based HAR tasks}.
For example, \textit{how to enable personalization without requiring extensive personal data or prolonged calibration}, 
and \textit{how to effectively apply federated learning to train a strong global model while allowing rapid specialization for individual users}.

\subsubsection{\textbf{Complex and Composite Activities}}

Existing HAR studies usually concentrate on atomic activities, where only a single action occurs at a time. 
However, real-world daily life frequently involves concurrent or overlapping activities, for example, walking while talking on the phone or cooking while watching television, as well as complex activities composed of temporally organized sub-actions. 
For instance, the activity of “making coffee” typically entails multiple interdependent steps.
Accurately recognizing such composite activities may require hierarchical modeling approaches \cite{lan2012social}, in which primitive actions are first detected and subsequently structured into higher-level sequences, or multi-label classification methods \cite{jethanandani2020multi} capable of capturing simultaneous activities. 
Another pertinent issue concerns activity intensity and quality. 
For example, differentiating vigorous from mild running or detecting improper exercise form is essential in fitness and rehabilitation contexts. 
Although early studies have addressed composite activities and intensity-aware recognition, a comprehensive solution remains elusive. 
This challenge is closely linked to context awareness, which requires understanding not only motion patterns but also cues such as time, location, and object interactions to resolve activity ambiguity. 
IMU signals alone are often insufficient, underscoring the importance of multi-modal sensing. 
While recent work has explored integrating supplementary sensors, more advanced modeling techniques are still needed to capture these nuanced aspects effectively.

\subsubsection{\textbf{Privacy and Security Concerns}} 

Privacy and security remain critical yet unresolved challenges in IMU-based HAR. 
While protecting user data motivates methods like federated learning and differential privacy, over-obfuscation can impair model performance. 
Balancing privacy and utility remains an open problem, as techniques such as noise injection often degrade accuracy. 
In addition, security concerns are also critical for HAR tasks. 
Adversaries could inject false sensor data to mislead HAR systems. 
For instance, simulating a fall to trigger false alarms. 
Strengthening HAR models against such attacks is an emerging priority and still underexplored.
Moreover, privacy and utility often conflict: while IMU data can reveal sensitive personal habits, it is also essential for health monitoring. 
This tension highlights the need for user-centric control that enables beneficial inferences while preventing unwanted disclosures.

\subsubsection{\textbf{Evaluation and Standardization}}

A broader community challenge in HAR is the lack of standardized evaluation protocols and common benchmarks. 
The UCI-HAR dataset \cite{anguita2013public} from 2012, still widely used as of 2024, reflects its longevity but also raises concerns of potential overfitting.
Moreover, evaluation should extend beyond classification accuracy to measure downstream impact. 
For example, if a HAR system supports lifestyle coaching, the true metric is improved health outcomes. Incorporating such holistic metrics remains challenging but essential for practical relevance. 
Community-driven efforts, such as challenge competitions, can promote progress. 
Past examples include the SHL transportation mode challenge \cite{wang2021locomotion} and the Kaggle Parkinson’s gait detection competition \cite{tlvmc-parkinsons-freezing-gait-prediction}, both of which encouraged testing under realistic conditions.

\subsection{Future Directions}

Looking ahead, multiple research avenues are expected to shape the future of IMU-based sensing and HAR. 
These include the integration of foundation models and large language models (LLMs) into HAR, the application of generative AI, the development of federated and continual learning ecosystems, and broader interdisciplinary collaboration.

\subsubsection{\textbf{Foundation Models and Large Language Models in HAR}}

Foundation models and LLMs are poised to play a transformative role in the future development of HAR systems. 
Their capacity for multimodal integration, contextual reasoning, and semantic understanding opens up several promising research directions.

\textbf{Training Foundation Models with Multimodal Sensor Data:}
Recent studies have begun exploring the design of foundation models tailored to HAR \cite{zhang2024unimts,Xu2021,xu2025lsm,narayanswamy2024scaling}. 
While current LLMs have demonstrated zero-shot capabilities on raw IMU data \cite{hargpt}, they are not optimized for processing sensor signals. 
A promising direction involves training foundation models on multimodal datasets that combine sensor data with auxiliary Modalities. 
For example, collecting wearable sensor streams alongside real-time verbal activity descriptions could enable models to learn semantically rich representations of human motion. 

\textbf{Grounding LLMs in Physical Sensor Data:}
A central challenge in applying LLMs to HAR is ensuring factual grounding, i.e., constraining models to produce physically plausible interpretations of sensor input. 
Approaches such as prompt engineering \cite{white2023prompt} may be employed to guide model reasoning with task-relevant statistics (e.g., “The variance of acceleration is X, typical of running”). 
Another promising avenue involves hybrid model architectures in which LLMs serve as reasoning agents or controllers that delegate feature extraction and classification to specialized, lightweight sensor models.
In addition, by reasoning in natural language, models could provide transparent justifications for their decisions. 
This could help mitigate the black-box nature of deep learning-based HAR systems and build user trust.

\textbf{Leveraging Contextual Reasoning of LLM in Ambiguous Scenarios:}
Ambiguity is common in IMU-based HAR, as different activities may produce similar sensor patterns 
(e.g., cycling versus driving on a bumpy road). 
When available, location data, e.g., from human GPS trajectory~\cite{guo2025leveraging} or indoor positioning systems~\cite{guo2022wepos}, can provide powerful priors. 
For example, activity patterns in a gym are likely exercise-related, while those in a kitchen may pertain to cooking. 
Integrating contextual signals such as location, time of day, and even calendar events (e.g., user schedules) enables more accurate inference. 
Some works assistants already implement rudimentary versions of this (e.g., commuting detection \cite{yu2020mobile}). 
The key research challenge lies in effectively fusing symbolic context with raw sensor analytics, such as location categories, time semantics, and known routines. 
One of the advantages of large language models lies in their reasoning capacity to incorporate contextual knowledge. 
Integrating contextual knowledge or a probabilistic graphical model into LLM may offer promising solutions for this type of contextual reasoning.

\textbf{Semantic and Human-Centric Activity Understanding with LLMs:}
Future HAR systems are expected to evolve from flat, unstructured activity labels toward semantically grounded and human-centric representations. 
Large language models (LLMs), with their ability to capture rich world knowledge and reason over context, 
offer unique opportunities to automatically construct and update semantic activity ontologies, 
understand relationships among activities, 
and generalize recognition to previously unseen or infrequent actions. 

Beyond identifying what a user is doing, LLM-enhanced HAR systems could infer why an activity is performed 
and even anticipate what might happen next, enabling proactive and personalized assistance. 
For example, recognizing that a user is putting on running shoes could trigger relevant health services before the activity begins. 
Combining wearable sensing with LLM-driven reasoning about user context, routines, and intentions 
paves the way for HAR systems that are not only accurate but also deeply human-centric, 
offering semantically meaningful, context-aware, and predictive insights.

\subsubsection{\textbf{Cross-Modal and Generative AI for Data Augmentation and Adaptation}}

Data limitations and distribution shifts remain major bottlenecks in IMU-based HAR. 
Leveraging cross-modal signals and generative AI offers new opportunities to enhance data diversity and model adaptability.

\textbf{Generative Models for Data Synthesis and Augmentation:}
Generative models (e.g., VAE \cite{kingma2013auto}, GAN \cite{goodfellow2020generative}, Flow matching \cite{lipman2022flow}, and Diffusion models \cite{ho2020denoising}) offer promising avenues for augmenting HAR datasets by synthesizing realistic sensor sequences. 
These models can simulate rare or hazardous scenarios (e.g., falling down stairs), which are difficult to capture from real subjects but critical for safety-critical applications such as elder care. 
Conditional generation, based on attributes like age or fitness level, may further address data imbalance and improve model generalization. 
Ensuring the fidelity and utility of synthetic data remains a key concern, requiring rigorous evaluation through metrics such as feature distribution similarity.
Future advances may integrate generative AI into physics-based simulation environments, enabling virtual agents to generate multimodal sensor data aligned with embodied AI objectives. 
In addition, achieving controllable and conditional generation of IMU sensor data for long-duration activity sequences with realistic transitions remains an open and active research direction.

\textbf{Generative Models for Domain Adaptation:}
Generative models offer powerful tools for mitigating domain shifts through style transfer, such as transforming smartphone sensor data to mimic smartwatch signals or adapting motion patterns across demographics (e.g., young vs. older adults). 
By explicitly modeling inter-domain differences, sensor inputs can be converted between domains, as demonstrated in early works like CrossHAR~\cite{crosshar}. 
Future approaches may adopt conditional diffusion models or other generative modeling approaches to perform such transformations implicitly, enabling training pipelines to leverage both real and domain-shifted variants for improved generalization. 
Cross-modal adaptation further expands this potential, enabling integration of novel sensors without extensive retraining. 
For instance, generative translation between EMG and IMU signals—given sufficient inter-modality correlation—could facilitate multi-sensor fusion. 
Existing works such as IMG2IMU \cite{Leng2024IMG2IMU} illustrate knowledge transfer across modalities, hinting at future exploration of modalities like textual descriptions or audio cues that also encode activity patterns. 
The long-term vision is a universal generative model of human motion, capable of producing diverse sensor outputs (e.g., accelerometer, gyroscope, video, RF signals) from high-level descriptions such as “a 70-year-old slowly climbing stairs while holding a handrail”. 
In the short term, specialized generative augmentation tools, e.g., inject noise or coordinate transformation, are expected to strengthen the robustness and generalizability of HAR models.

\subsubsection{\textbf{Federated and Continual Learning Ecosystems}}

Ensuring privacy, personalization, and adaptability in IMU-based HAR systems calls for learning paradigms that operate across distributed users and evolving contexts.

\textbf{Federated Model Ecosystem:}
Federated learning is expected to evolve into more complex ecosystems, moving beyond the traditional server–client paradigm. 
One emerging direction is federated continual learning \cite{yoon2021federated}, wherein models incrementally improve post-deployment by learning from new data in a distributed manner. 
This establishes a virtuous cycle: each user's device contributes to model refinement, benefiting all users through subsequent updates. 
A key direction is mitigating catastrophic forgetting—the loss of previously acquired knowledge—while integrating new patterns. 
Techniques such as replay buffers and model regularization are under active investigation within the federated learning framework.

\textbf{Personal federated models:}
The concept of personal federated models also offers a compelling future trajectory. 
Instead of restarting personalization when switching devices (e.g., from a Fitbit to an Apple Watch), users could carry their personalized model across platforms. 
With standardized and privacy-preserving (e.g., encrypted) model formats, one could export a model from one device and seamlessly import it into another with minimal recalibration. Achieving this would require cross-platform model interoperability \cite{soursos2016towards}, potentially enabled by open-source standards. 

\subsubsection{\textbf{Energy Efficiency and Resource-constrained Inference}} 

As IMU-based HAR systems move toward real-world deployment, energy efficiency and the ability to operate under limited computational resources (e.g., smartphones and smartwatches) have become increasingly critical.

\textbf{Energy Efficiency:} For wearable devices to continuously monitor activities, energy efficiency is a critical concern. 
High-frequency sensor sampling and intensive computation can rapidly deplete battery life. 
A key open challenge lies in adaptive sensing: dynamically adjusting sensor sampling rates or duty cycles based on detected activity or contextual cues to conserve energy without sacrificing detection of important events. 
Some prior work has explored context-aware duty cycling \cite{jemili2017context}, but integrating such adaptive schemes with contemporary HAR models remains an open problem.

\textbf{Model Compression and Resource-constrained Inference:} 
On the computational aspect, executing deep neural networks on devices with limited computational resources poses substantial difficulties. 
Approaches such as TinyML address this challenge but often incur accuracy trade-offs \cite{daghero2022human}.
Thus, a major challenge is compressing HAR models (through quantization, pruning, or related techniques) to minimize accuracy loss while enabling efficient on-device inference. 
Another critical area requiring further research is hardware acceleration. 
Although recent IMU chips support on-chip early feature extraction and dedicated neural network accelerators can execute lightweight models at ultra-low power, exploiting these hardware capabilities through algorithm–hardware co-design remains an open challenge that demands greater effort.
\section{Conclusion}
\label{sec:conclusion}

In this paper, we present a comprehensive survey of generalizable HAR utilizing IMU sensors from mobile and wearable devices. 
We first outline the overall framework and existing generalization-oriented training settings for IMU-based HAR. 
We then review, summarize, and categorize existing studies from multiple perspectives, including datasets and benchmarks, data-centric methodologies, model-centric methodologies, and applications. 
These investigations demonstrate that generalization is critical for IMU-based HAR and has broad scientific and real-world impacts. 
Finally, we discuss key challenges, open problems, and future directions from the perspectives of data preprocessing, model training, evaluation, and deployment. 
We hope this work will help researchers and the broader community better understand, apply, and advance generalizable IMU-based HAR research, thereby promoting the practical deployment of HAR models.


\bibliographystyle{ACM-Reference-Format}
\bibliography{survey}

\end{document}